\documentclass[preprints,article,accept,moreauthors,unicode]{Definitions/mdpi} 

\firstpage{1} 
\pubvolume{1}
\issuenum{1}
\articlenumber{0}
\pubyear{2026}
\copyrightyear{2026}
\datereceived{ } 
\daterevised{ } 
\dateaccepted{ } 
\datepublished{ } 

\usepackage{braket}
\usepackage{qcircuit}
\usepackage{amssymb} 
\usepackage{tabularx}
\usepackage{array}
\usepackage{algorithmic}
\usepackage{algorithm}
\usepackage{mathtools}
\definecolor{groupgray}{RGB}{238,238,238} 
\usepackage{caption}
\usepackage{graphicx}
\usepackage[table]{xcolor}

\newcommand{\cmark}{\checkmark}
\newcommand{\xmark}{$\times$}

\Title{Resource-Efficient Quantum Reinforcement Learning on NISQ Devices with Dynamic-Circuit Qubit Reuse and Grover-Based Trajectory Search}

\Author{Thet Htar Su $^{1,}$*\orcidA{}, Shaswot Shresthamali $^{2}$ and Masaaki Kondo $^{1,3}$}

\AuthorNames{Thet Htar Su, Shaswot Shresthamali and Masaaki Kondo}

\address{%
$^{1}$ \quad Graduate School of Science and Technology, Keio University, Yokohama, Kanagawa 223-8522, Japan\\
$^{2}$ \quad Graduate School of Information Science and Electrical Engineering, Kyushu University, Fukuoka 819-0395, Japan\\
$^{3}$ \quad RIKEN Center for Computational Science, Kobe 650-0047, Japan}

\corres{Correspondence: thethtarsu@acsl.ics.keio.ac.jp}

\abstract{
A resource-efficient quantum reinforcement learning framework for finite-horizon quantum Markov decision processes is presented, employing dynamic-circuit qubit reuse to reduce the interaction-workspace requirements of multi-step execution. The framework integrates quantum state-transition and reward encoding, dynamic-circuit execution, quantum return evaluation, and Grover-based trajectory search. Its central contribution is a dynamic execution model in which a fixed interaction workspace is reused across sequential agent–environment interactions through mid-circuit measurement, qubit reset, and next-state propagation. For the considered benchmark, this approach reduces the interaction-workspace requirement from 7\(T\) qubits under static circuit to a constant seven qubits, independent of the interaction horizon \(T\). This corresponds to a 66\% reduction for the three-step benchmark and transforms the interaction-workspace complexity from \(O(T)\) to \(O(1)\). Trajectory returns are evaluated using reversible quantum arithmetic, and Grover’s algorithm identifies the optimal trajectories, so that return evaluation and trajectory search are performed within a unified quantum-native pipeline. Ideal simulations demonstrate that the dynamic implementation preserves the valid trajectory structure of the static formulation, with quantitative comparisons assessing the agreement between their trajectory distributions. Finally, execution on an IBM Heron-class processor demonstrates the feasibility of the proposed dynamic-circuit architecture on noisy intermediate-scale quantum hardware.
}


\begin{document}

\section{Introduction} \label{sec:introduction}

Classical reinforcement learning (RL) has achieved remarkable success in sequential decision-making tasks. However, it often suffers from slow convergence and substantial computational demands. Inefficient exploration and high sample complexity remain persistent challenges, particularly in large or complex environments, where extensive interactions are required to identify optimal policies~\cite{challenges_RL}. To address these limitations, quantum reinforcement learning (QRL) has emerged as a promising paradigm to investigate how quantum computation may support reinforcement-learning processes~\cite{ chen2024}. By exploiting quantum-mechanical resources such as superposition and amplitude amplification, QRL has been studied for potential improvements in exploration, sampling, and decision-making efficiency~\cite{qml2025}. Early research explored hybrid quantum--classical approaches, in which quantum circuits are integrated into selected components of the RL pipeline, as well as quantum-inspired policy-optimization algorithms~\cite{meyer2022surveyRL}. Although such hybrid methods provide valuable insights, they may introduce additional overhead associated with repeated classical--quantum data exchange.

Therefore, researchers have also investigated fully quantum RL formulations in which the agent--environment interaction and associated computation are represented within quantum circuits~\cite{wiedemann2023,thet2024}. Earlier work~\cite{wiedemann2023} established proof-of-concept formulations, whereas a subsequent study~\cite{thet2024} exposed an important resource limitation in multi-step implementations. In the static quantum Markov decision process (QMDP) construction considered in~\cite{thet2024}, a four-state, two-action interaction required seven qubits, and extending it over \(T\) interaction steps required a separate set of interaction registers for each step, resulting in a \(7T\)-qubit interaction workspace. Such static circuit unrolling rapidly increases qubit resource requirements and becomes increasingly difficult to realize on present-day quantum processors, where the number of practically usable qubits is constrained by noise and operational imperfections~\cite{NISQ2018,QRCC2024}. This limitation arises from the execution architecture used to represent successive agent--environment interactions rather than from the state and action encoding alone. Under static unrolling, increasing the interaction horizon requires repeated allocation of the state, action, next-state, and reward registers, so the width of the interaction workspace grows linearly with the horizon. Richer agent–environment modeling has therefore been achievable only at the cost of linear qubit scaling, which restricts the practical implementation of longer sequential interactions on near-term hardware.

Overcoming this limitation requires decoupling the interaction depth from the physical register allocation while preserving the trajectory evolution of the finite-horizon process. Dynamic circuits provide this capability through mid-circuit measurement, reset, and qubit reuse. Rather than allocating independent interaction registers to every time step, the same state, action, next-state, and reward registers can be reused across successive interactions. At each step, the state--action transition and reward generation are evaluated in quantum registers, followed by measurement, sequential next-state propagation, reset, and reuse for the subsequent interaction. This approach reduces the dependence of the reusable interaction workspace on the interaction horizon while retaining the sequential structure of the finite-horizon process.

Building on this principle, this study introduces a resource-efficient QRL architecture for finite-horizon QMDP. Unlike the static construction in~\cite{thet2024}, which allocates a separate set of interaction registers for each time step, the proposed architecture employs dynamic-circuit qubit reuse to execute successive agent--environment interactions using a fixed interaction workspace. The contribution is primarily architectural and resource-oriented: it reformulates the execution of multi-step QMDPs without changing the underlying decision-making problem. For the benchmark considered, the interaction-workspace requirement is reduced from $7T$ qubits to a fixed workspace of seven reusable qubits, while additional registers required for reward retention and return evaluation are accounted for separately. Building on this execution model, quantum return evaluation and Grover-based trajectory search are integrated into the workflow. The specific contributions are as follows:
\begin{enumerate}
    \item Resource-efficient dynamic execution of multi-step QMDP: A dynamic-circuit architecture is developed to overcome the linear growth of the interaction workspace associated with static circuit construction in~\cite{thet2024}. The state, action, next-state, and reward registers are reused across successive interactions through mid-circuit measurement, sequential next-state propagation, reset, and re-initialization. This approach reduces the interaction-workspace complexity from $O(T)$ to $O(1)$ with respect to the interaction horizon, while additional registers required for reward retention and return evaluation are accounted for separately.
    \item Integration of quantum return evaluation and Grover-based trajectory search: Trajectory returns are evaluated using reversible quantum arithmetic, and Grover-based search is incorporated into the dynamic-circuit workflow to identify optimal trajectories. This combines multi-step trajectory generation, return evaluation, and optimal-trajectory search within a unified quantum execution framework.
    \item Experimental validation on NISQ hardware: The proposed dynamic-circuit QRL architecture is implemented and evaluated on an IBM Heron-class quantum processor. The hardware experiments demonstrate its practical feasibility and assess its performance under realistic hardware noise.
\end{enumerate}

Together, these contributions establish a resource-efficient approach to multi-step quantum agent--environment interaction and optimal trajectory search, providing a practical framework for finite-horizon QRL. The remainder of this paper is organized as follows: Section~\ref{sec:related_work} reviews the related work; Section~\ref{sec:background} introduces the relevant concepts; Section~\ref{sec:method} presents the proposed architecture; Section~\ref{sec:experiments_results} reports the experimental results; Section~\ref{sec:discussion} discusses the findings; and Section~\ref{sec:conclusion} concludes the paper.

\section{Related Work}
\label{sec:related_work}
QRL has been developed through two main approaches: hybrid quantum–classical methods and fully quantum methods. The following review summarizes the key developments in both approaches and motivates the need to address the resource limitations considered in this study.

\subsection{Hybrid Quantum-Classical RL}
Hybrid quantum–classical RL includes quantum-inspired RL (QiRL), variational quantum circuit (VQC)-based RL, and RL methods that leverage quantum subroutines.

QiRL represents the initial integration of quantum computing (QC) and RL, aiming to overcome classical limitations such as slow learning, the exploration–exploitation trade-off, and high computational costs~\cite{meyer2022surveyRL}. By leveraging quantum principles such as superposition and amplitude amplification, researchers have advanced the field from representing actions and policies within the quantum domain~\cite{Don08,Don06,Chun08,Chen06,Don12,Ganger2019,Cho2022} to embedding experiences in deep RL~\cite{Wei22}. These theoretical developments have been validated in diverse domains, including robotics~\cite{Don06,Don12}, unmanned aerial vehicles (UAVs)~\cite{Li1994}, and clinical decision-making~\cite{Li2020,Niraula2021}, demonstrating the potential of QC to accelerate learning and reduce the computational overhead in real-world applications~\cite{meyer2022surveyRL}.

VQC-based RL employs variational quantum circuits (VQCs) as replacements for classical neural networks in deep RL to address the limitations of the QiRL approach in high-dimensional problems~\cite{Sequeira2022}. VQCs were initially used as function approximators in double deep Q-learning for simple environments such as Frozen Lake~\cite{Chen20}, and were later applied to complex tasks such as Cart Pole~\cite{LockwoodSi20}, Atari games~\cite{lockwood21a}, and continuous action spaces~\cite{Wu_2020}. Some studies~\cite{Skolik2022} have shown that performance depends on architecture and hyperparameter choices, enabling significant model simplification compared to classical methods. Beyond value-based methods, VQCs have also been applied in policy-based algorithms such as REINFORCE~\cite{Sequeira2022,jerbi_2022} and Actor–Critic~\cite{Kwak21,Lan2021}. Studies have reported advantages including faster convergence, reduced memory consumption, and fewer parameters than classical RL in complex environments.

RL with quantum subroutines integrates quantum algorithms into the classical RL workflow. Notable examples include quantum value iteration, which combines quantum mean estimation and maximum finding to reduce sample complexity~\cite{wang2021}, and quantum policy iteration, which alternates between quantum policy evaluation with quantum linear system solvers and classical policy improvement~\cite{Cherrat2023}. These approaches offer potential computational advantages over their classical counterparts under specific assumptions.

\subsection{Full Quantum Reinforcement Learning}
Earlier work on full QRL~\cite{wiedemann2023} introduced a quantum policy iteration framework that leveraged quantum superposition for trajectory generation, amplitude estimation for policy evaluation, and Grover’s algorithm for policy improvement. This approach substantially reduced the sample complexity relative to classical RL. A notable recent contribution~\cite{thet2024} advanced this direction by encoding the components of a finite-state Markov decision process, including state representation, action selection, state transitions, and rewards, within a quantum-circuit formulation. Quantum arithmetic was used to evaluate trajectory returns, followed by Grover-based search to identify the high-return trajectories. That study demonstrated the feasibility of combining quantum trajectory generation, return evaluation, and trajectory search in a multi-state, multi-step setting.

However, this prior work~\cite{thet2024} incurred a substantial quantum-resource overhead because the interaction registers were statically replicated across multiple time steps. A single agent--environment interaction required seven qubits to encode the state, action, next state, and reward registers, and extending the same construction to \(T\) sequential interactions therefore required a \(7T\)-qubit interaction workspace. This linear growth in the interaction-workspace width increases the physical-qubit requirement with the interaction horizon and constrains the implementation of multi-step QRL on near-term quantum processors, where usable qubit resources are limited by hardware availability, connectivity, and operational constraints. 

Dynamic-circuit techniques have emerged as a promising approach to improve qubit resource efficiency through qubit reuse. CaQR~\cite{CaQr2023} leveraged dynamic circuits to automatically identify qubit-reuse opportunities and reduce circuit width while preserving computational correctness. DeCross et al.~\cite{DeCross2023} demonstrated the practical viability of qubit recycling by compressing an 80-qubit QAOA circuit to 20 qubits, while QRCC~\cite{QRCC2024} combined qubit reuse with circuit cutting to reduce physical-qubit requirements and classical post-processing costs. Collectively, these studies demonstrate the potential of dynamic-circuit-based qubit reuse for reducing hardware-width requirements.

Motivated by these advances, this study reformulates the static QMDP execution model using dynamic-circuit-based qubit reuse. Mid-circuit measurement, reset, and sequential next-state propagation allow the same state, action, next-state, and reward registers to be reused across successive interaction steps rather than replicated at each step. As a result, the reusable interaction-workspace width is decoupled from the interaction horizon, while the sequential structure of the finite-horizon decision process is retained. The proposed architecture further integrates quantum return evaluation and Grover-based trajectory search within the same dynamic-circuit workflow. Together, these advances provide a resource-efficient QRL framework that reduces the physical-qubit overhead associated with static circuit unrolling and improves the feasibility of finite-horizon implementations on the noisy intermediate-scale quantum (NISQ) hardware.

\section{Background}\label{sec:background}

\subsection{Reinforcement Learning}
Reinforcement learning is a framework for sequential decision-making, in which an agent interacts with an environment to learn a policy that maximizes cumulative rewards through trial and error learning~\cite{sutton2018reinforcementintroduction,graesser2019deepfoundations,Goodfellow-et-al-2016}. RL has been successfully applied to complex decision-making problems in areas such as autonomous driving~\cite{shalevshwartz2016}, robotics~\cite{Rl_robotics_a_survey}, and game playing~\cite{Silver2017,Noam_Superhuman_AI}.

RL is commonly formulated as a Markov decision process (MDP), defined by the state space $S$, action space $A$, reward function $\textit{R}(s,a):=\mathbb{E}[r_t|s_t=s,a_t=a]$, and transition function $P(s_{t+1}|s_t,a_t)$~\cite{meyer2022surveyRL}. At each time step $t$, the agent observes a state $s_t\in S$, selects an action $a_t\in A$, receives a reward $r_t$, and transitions to the next state $s_{t+1}$, generating a trajectory
$s_0,a_0,r_0,s_1,a_1,r_1,\ldots,s_T$~\cite{sutton2018reinforcementintroduction}. The objective is to learn a policy $\pi$ that maximizes the expected cumulative reward~\cite{Plaat_2022}. This MDP formulation provides the classical foundation for the QRL framework considered in this study.

\subsection{Quantum Computing}
Quantum computing exploits quantum-mechanical principles to process information using quantum bits (qubits), which can exist in superpositions of computational basis states~\cite{quantum_gentleintro}. Quantum computations typically involve state initialization, unitary operations, and measurement. Quantum gates transform the quantum state while preserving normalization, whereas measurement produces classical outcomes according to the resulting probability distribution~\cite{nielsen2010quantum}.
Superposition enables quantum systems to represent multiple computational basis states within a single quantum state, while quantum interference can increase the probability of desirable outcomes and suppress others. These properties underlie quantum algorithms such as Shor's factoring algorithm~\cite{Shor1994Algorithms,Shor1994Algorithms_97,Shor_1996} and Grover's search algorithm~\cite{groveralgorithm}. In this study, Grover's algorithm is employed as a quantum subroutine for searching the trajectory space and identifying optimal candidates.

Modern quantum hardware also supports dynamic circuits, which incorporate mid-circuit measurement, classical feed-forward, and conditional quantum operations during execution~\cite{johnson2022,errorcorrection2022,digiq2022,surface2022,qdb2019,runtime2020}. Mid-circuit measurement can extract information from selected qubits while allowing the remaining qubits to continue evolving, and the measured qubits can be reset and reused within the same circuit. Consequently, dynamic circuits can reduce the number of physical qubits required for multi-step computations and provide a resource-efficient alternative to statically unrolled circuits~\cite{CaQr2023}. These capabilities are particularly relevant to the proposed QMDP architecture, which reuses a fixed set of quantum registers across successive agent--environment interactions.

\section{Method}\label{sec:method}
\subsection{Overview of the Proposed QRL Architecture}
The proposed QRL framework integrates quantum processing with dynamic-circuit execution to realize multi-step agent--environment interactions while enabling the systematic reuse of the interaction registers. At each interaction step, the state, action, next-state, and reward registers undergo quantum processing through unitary operations, including superposition preparation, transition encoding, and reward generation. The proposed architecture does not maintain a coherent superposition of complete multi-step trajectories across the full interaction horizon; instead, coherent quantum processing is localized to each individual agent--environment interaction.

Before the interaction registers are measured, the reward information required for the subsequent return evaluation is retained in the dedicated ancillary quantum registers. The state, action, next-state, and reward registers are then measured, and the corresponding outcomes are stored in classical registers for trajectory reconstruction. The measurement projects the interaction registers onto a sampled branch; consequently, the retained ancillary reward registers correspond to the sampled interaction history rather than preserving a coherent superposition over all the preceding trajectory branches. Following each mid-circuit measurement, the state, action, and reward registers are reset. The next-state register remains in its post-measurement computational-basis state and is then used as the control of CNOT operations to initialize the state register for the subsequent interaction, after which the next-state register is reset for reuse. This procedure allows the same interaction registers to be reused across successive agent–environment interactions while preserving the state transition between steps. 

After the final interaction, the retained reward registers together with the final-step reward register are used to construct the quantum return register through reversible quantum arithmetic. Grover-based trajectory search is then applied before the final measurement, using the retained reward and return information together with registers associated with the final interaction rather than a coherent quantum register containing the complete multi-step state--action trajectory. Earlier interaction outcomes remain available through their classical measurement records and are combined with the final measurement outcomes to reconstruct the complete trajectory.

Therefore, the key methodological contribution is a dynamic execution model for a multi-step QMDP in which the same interaction registers are reused across sequential interaction steps without maintaining trajectory-wide quantum coherence. Quantum operations are performed within each agent--environment interaction and subsequently on the retained reward and return registers, while progression between interaction steps is realized through measurement, reset, and next-state propagation. Although the dynamic QMDP and Grover-based trajectory search are implemented within the same dynamic quantum circuit, the overall computation does not constitute a single globally coherent unitary evolution across the complete interaction horizon because of the intermediate measurement and reset operations. This architecture reduces the horizon-dependent growth of the reusable interaction workspace while retaining the information required for trajectory reconstruction, quantum return evaluation, and optimal trajectory search. The following subsections describe the QMDP encoding, register allocation, dynamic-circuit-based qubit reuse, return calculation, and Grover-based trajectory search in detail.

\subsection{Quantum Markov Decision Process}
In RL, an agent interacts with an environment across discrete time steps. At each step, the agent observes a state, selects an action, and the environment responds with a new state and an associated reward. The probability distribution over the next state is determined by the state transition function, while the reward function assigns numerical values to transitions, reflecting the outcome of a given state--action pair. In the classical setting, states and actions are represented as finite sets, the transition function is described by a stochastic matrix, and rewards are specified numerically. In the quantum formulation considered here, these MDP components are represented by quantum registers and operations. The present work focuses on a stochastic environment with four states $(s_0,s_1,s_2,s_3)$ and two actions $(a_0,a_1)$ as a benchmark instance of the general construction described below.

\subsubsection{Quantum State and Quantum Action}
In the proposed QMDP, discrete states and actions are represented using computational-basis encoding. For arbitrary finite state and action spaces, \(\mathcal{S}=\{s_0,\ldots,s_{|\mathcal S|-1}\}\) and \(\mathcal{A}=\{a_0,\ldots,a_{|\mathcal A|-1}\}\), the corresponding register widths are \(n_s=\left\lceil\log_2|\mathcal S|\right\rceil\) and \(n_a=\left\lceil\log_2|\mathcal A|\right\rceil\).
Each classical state \(s_i\) and action \(a_j\) is mapped to the computational-basis states \(\ket{i}_S\) and \(\ket{j}_A\), respectively. If \(|\mathcal S|\) or \(|\mathcal A|\) is not a power of two, the additional computational-basis states do not correspond to valid states or actions and are therefore excluded from the valid encoding space.
For the benchmark considered in this study, $\mathcal{S}=\{s_0,s_1,s_2,s_3\}$ and $\mathcal{A}=\{a_0,a_1\}$, requiring two state qubits and one action qubit. The state encoding is $ s_0 = \ket{00}, s_1 = \ket{01}, s_2 = \ket{10}, s_3 = \ket{11},$
while the action encoding is $a_0  = \ket{0}, a_1 = \ket{1}.$

At the initial interaction step, the state register is initialized in a uniform superposition over the valid benchmark states using Hadamard gates as follows in Equation (\ref{eq:method_one}):
\begin{equation}
   H^{\otimes 2}(\ket{0_s} \otimes \ket{0_s}) = \sum_{n = 0}^{N-1} c_n \ket{s_n}, \quad  \text{with }  c_n = \frac{1}{\sqrt{N}},
\label{eq:method_one}
\end{equation}
with normalization \(\sum_{n=0}^{N-1} |c_n|^2 = 1.\)
For $N=4$,
\[ H^{\otimes 2}(\ket{0_s} \otimes \ket{0_s}) = \frac{1}{2} \ket{00} + \frac{1}{2} \ket{01} + \frac{1}{2} \ket{10} + \frac{1}{2} \ket{11}.\] 
This state satisfies the normalization condition $\sum_{n=0}^{3}|c_n|^2=1$ with $c_n=1/2$. 
Subsequent interaction steps do not prepare this uniform state superposition; instead, the state register is initialized from the next-state register of the preceding interaction through the state-propagation procedure described later.

Similarly, the action qubit is prepared at each interaction step in a uniform superposition using Hadamard operation (Equation (\ref{eq:method_three})):
\begin{equation}
    H\ket{0_a} = \sum_{n=0}^{1} c_n \ket{a_n}, \quad c_n = \frac{1}{\sqrt{2}},
\label{eq:method_three}
\end{equation}
which gives
\[ H\ket{0_a} = \frac{1}{\sqrt{2}}\ket{0} + \frac{1}{\sqrt{2}}\ket{1}.\]

In the present implementation, the action register does not represent a trainable parameterized policy, and no iterative policy-update rule is applied during the interaction sequence. Instead, the uniform action superposition provides candidate actions for finite-horizon trajectory generation. Trajectory performance is evaluated through the accumulated return, and the subsequent trajectory-search stage identifies optimal trajectories. Accordingly, the proposed procedure performs finite-horizon trajectory generation and identification rather than conventional iterative policy learning.

\subsubsection{Quantum State Transition Function}
Once the state and action registers have been prepared, the MDP transition function is encoded in the quantum circuit. Classically, $P(s'|s,a)$ specifies the probability of transition from the current state $s$ to the next state ${s'}$ under action $a$. In the proposed QMDP, these transition probabilities are encoded in the amplitudes of the next-state register, such that their squared magnitudes reproduce the corresponding classical transition probabilities.

For a state--action pair $(s_i,a_j)$ with $K$ nonzero successor states and probabilities \(p_k=P(s'_k|s_i,a_j)\), the transition operator prepares
\[U_P\ket{s_i}\ket{a_j}\ket{0}^{\otimes n_s}=\ket{s_i}\ket{a_j} \sum_{k=0}^{K-1} \sqrt{p_k}\ket{s'_k},
\quad  \text{where} \sum_{k=0}^{K-1}p_k=1. \]
Consequently, measurement of the next-state register yields $s'_k$ with probability $p_k$.

In the present implementation, the conditional next-state preparation is realized using controlled $R_y$ rotations. The choice of $R_y$ follows from
\[ R_y(\theta)\ket{0} = \cos\left(\frac{\theta}{2}\right)\ket{0} + \sin\left(\frac{\theta}{2}\right)\ket{1}. \]
For a binary conditional branch in which the probability of the $\ket{1}$ branch is $p$, the required rotation angle is \(\theta=2\arcsin(\sqrt{p}).\)
For multiple successor states, the next-state distribution is constructed recursively using conditional branch probabilities associated with successive next-state qubits. This corresponds directly to the conditional state-preparation procedure implemented in the transition circuit.

In the four-state benchmark, each stochastic state--action pair has at most two nonzero successor states, while a deterministic transition contains a single successor with probability one. More generally, distributions with additional successor states can be prepared using the same conditional-rotation construction. Alternative state-preparation approaches, including multiplexed \(R_y\) rotations, may also be employed for larger transition models.
The rotations are conditioned on the computational-basis values of the current state and action registers such that the appropriate transition distribution is applied only to the corresponding state--action pair. For a particular pair \((\ket{s^*},\ket{a^*})\),
\begin{equation}
    CR_y(\theta) (\ket{s} \ket{a} \ket{0_{s'}}) =
    \begin{cases}
        \ket{s} \ket{a} R_y(\theta) \ket{0_{s'}} & \text{if} \ket{s}=\ket{s^*},\\ & \hspace{7pt} \ket{a}=\ket{a^*}, \\   
        \ket{s} \ket{a} \ket{0_{s'}}  & \text{otherwise}.
    \end{cases}
\label{eq:method_five}
\end{equation}

The cost of transition-state preparation increases with the number of state--action conditions and nonzero successor transitions. Let \(N_{\mathrm{succ}}\) denote the total number of nonzero successor transitions over all state--action pairs. For the controlled-\(R_y\) construction used here, the number of transition-encoding controlled rotations is upper-bounded by \(O(n_sN_{\mathrm{succ}})\) per interaction step and by \(O(Tn_sN_{\mathrm{succ}})\) over a horizon \(T\). The rotations are conditioned on the \(n_s+n_a\) state--action qubits, with additional controls on previously prepared next-state qubits where required by the conditional state-preparation sequence; consequently, decomposition into the native gate set becomes more costly as the register width and transition branching increase. The multi-controlled \(R_y\) operations in the present implementation do not require additional logical auxiliary qubits. The accuracy of the encoded transition distribution depends on the precision of the rotation synthesis and hardware errors. For larger transition models, alternative state-preparation approaches, including multiplexed \(R_y\)-rotation schemes, may offer different depth--accuracy trade-offs.

\subsubsection{Quantum Reward Function}
In a classical MDP, the reward function assigns a numerical value to a transition according to the underlying reward model. In the proposed QMDP, discrete reward values are represented using computational-basis encoding. 
For a finite reward set \(\mathcal{R}\), the reward-register width is 
\(n_r=\left\lceil\log_2|\mathcal{R}|\right\rceil,\)
with each reward value mapped to a corresponding computational-basis bit string.

More generally, the reward operation can be expressed as the reversible mapping
\[
\ket{s}\ket{a}\ket{s'}\ket{0}^{\otimes n_r}
\longrightarrow
\ket{s}\ket{a}\ket{s'}\ket{r(s,a,s')},
\]
where the reward register stores the value associated with the corresponding transition. In the present benchmark, the reward depends on the next-state encoding and is implemented using CNOT operations from the next-state register to the corresponding reward qubits. Thus, the CNOT-based reward circuit is a problem-specific realization of the general discrete reward mapping defined above.

\subsubsection{Single Agent---Environment Interaction}
A single agent--environment interaction within the QMDP is represented by quantum operations acting on the current-state, action, next-state, and reward registers. The interaction begins with the current state \(\ket{s}\). The action register is prepared as described above, after which the transition operation maps the current state and action to a superposition of possible next states \(\ket{s'}\), with amplitudes determined by the corresponding transition probabilities. The reward operation then encodes the associated reward in the reward register \(\ket{r}\).

Accordingly, before measurement, a single interaction can be represented by a joint quantum state of the form
\[
\ket{\phi}
=
\sum_{s\in\mathcal S}
\sum_{a\in\mathcal A}
\sum_{s'\in\mathcal S}
\sum_{r\in\mathcal R}
c_{s,a,s',r}
\ket{s}\ket{a}\ket{s'}\ket{r},
\]
where \(c_{s,a,s',r}\) denotes the amplitude associated with the corresponding state--action--next-state--reward configuration. The probability of obtaining a particular configuration upon measurement is
\(\left|c_{s,a,s',r}\right|^2\).

\subsection{Quantum Registers and Classical Storage}
The proposed algorithm employs a reusable set of interaction registers together with dedicated quantum registers for retained rewards and return evaluation, and a classical register for storing measurement outcomes. Table~\ref{table:quantumregister} summarizes the quantum and classical storage used in the implementation.

\begin{table}[H]
\scriptsize
\caption{Quantum registers and classical storage used in the proposed QMDP implementation.}
\label{table:quantumregister}
\renewcommand{\arraystretch}{1.15}

\begin{tabularx}{\textwidth}{
>{\centering\arraybackslash}p{1cm}
>{\centering\arraybackslash}p{1.1cm}
>{\centering\arraybackslash}p{2.2cm}
>{\raggedright\arraybackslash}X
>{\centering\arraybackslash}p{1.6cm}
}
\toprule
\textbf{Register name} &
\textbf{Type} &
\centering \textbf{Width (qubits/bits)} &
\centering \textbf{Stored information} &
\textbf{Reused across time steps} \\
\midrule
\texttt{qState} & Quantum & $n_s$ &
Encodes the current environment state & \cmark \\
\texttt{qAction} & Quantum & $n_a$ &
Encodes the candidate action; prepared in uniform superposition at each interaction step & \cmark \\
\texttt{qNextState} & Quantum & $n_s$ &
Encodes the next state generated by the environment transition & \cmark \\
\texttt{qReward} & Quantum & $n_r$ &
Encodes the reward generated at the current interaction step & \cmark \\
\texttt{qAnci[$t$]} & Quantum & $n_r$ per retained step &
Retains reward information from a preceding interaction step for subsequent return calculation and Grover-based trajectory search & \xmark \\
\texttt{qReturn} & Quantum & $n_g$ &
Encodes the trajectory return constructed from the retained earlier rewards and the final-step reward & \xmark \\
\texttt{cReg} & Classical & $T(2n_s+n_a+n_r)+n_g$ &
Stores the measured interaction outcomes across all time steps and the final return measurement for trajectory reconstruction & N/A \\
\bottomrule
\end{tabularx}
\end{table}

At each interaction step $t$, the quantum registers \texttt{qState}, \texttt{qAction}, \texttt{qNextState}, and \texttt{qReward} jointly encode a single agent--environment interaction of the QMDP. The measurement outcomes from these registers are stored in designated portions of the classical register \texttt{cReg}, which preserves the interaction records across time steps for subsequent trajectory reconstruction. 
Dedicated ancillary quantum registers retain the reward information generated at the preceding interaction steps before the corresponding interaction registers are measured and reset. These ancillary reward registers are not reset during the intermediate interaction cycles and are subsequently used in the quantum return calculation.

The quantum return register \texttt{qReturn} is constructed after the final interaction using the retained rewards from the preceding steps together with the final-step reward. The register widths $n_s$, $n_a$, $n_r$, and $n_g$ depend on the cardinality of the state space, action space, reward encoding, and representable return range, respectively. For the benchmark considered in this study, $n_s=2$, $n_a=1$, $n_r=2$, and $n_g=4$.

The reusable interaction workspace consists of \texttt{qState}, \texttt{qAction}, \texttt{qNextState}, and \texttt{qReward}, requiring \(2n_s+n_a+n_r\) qubits. These registers are measured, reset, and reused across the sequential interaction steps. The retained reward registers and \texttt{qReturn} are separate from this reusable interaction workspace and remain available for return evaluation and the subsequent Grover-based trajectory search. Accordingly, the constant-width qubit-reuse claim applies specifically to the interaction workspace, whereas the total quantum-resource requirement also includes the retained reward registers and the return register.

For a horizon of \(T\) steps, the total quantum width of the present implementation is
\( Q_{\mathrm{total}}=(2n_s+n_a+n_r)+(T-1)n_r+n_g,\)
where \(2n_s+n_a+n_r\) corresponds to the reusable interaction workspace, \((T-1)n_r\) accounts for the retained reward registers from the preceding interaction steps, and \(n_g\) is the width of the quantum return register. For the three-step benchmark, this gives \(7+4+4=15\) quantum qubits in total.

The classical register stores the measured interaction outcomes across all time steps together with the final return measurement and is used for trajectory reconstruction. For a horizon of \(T\) interaction steps, the required classical storage is \(T(2n_s+n_a+n_r)+n_g\) bits, where the first term stores the interaction records and the additional \(n_g\) bits store the final return measurement.

\subsection{Multiple Agent-environment Interactions with Dynamic Circuit Capability}
Extending QMDP to model sequential agent--environment interactions over a finite horizon $T$ requires not only a consistent quantum representation of state, action, next state, and reward at each time step, but also an execution model that remains feasible under realistic hardware constraints. A central contribution of this study is the realization of multi-step agent--environment interaction using a reusable interaction workspace, enabled by dynamic-circuit capabilities---specifically, mid-circuit measurement, reset, and next-state propagation. This design reduces the horizon-dependent qubit requirement of the interaction workspace and provides a resource-efficient execution model for multi-step QMDP.
The proposed architecture implements an agent--environment interaction loop as a sequential dynamic-circuit execution.
At each time step, quantum coherence is exploited to evaluate the state--action transition and reward generation in superposition, after which a projective measurement samples a concrete interaction outcome. For the intermediate interaction steps, the reward information required for the subsequent return evaluation is transferred to a dedicated ancillary register before measurement. The interaction registers are then measured and reused, while the next-state register provides the state information required to initialize the subsequent interaction. Algorithm~\ref{alg:dynamic_qmdp} formalizes the dynamic execution model for a $T$-step interaction.

Within this dynamic multi-step QMDP formulation, the circuit design must satisfy three core requirements at every intermediate time step: preserving the sampled interaction data, resetting the reusable qubits to the ground state $\ket{0}$, and propagating the environment state to the next time step. In particular, after measurement, the next-state register remains in its post-measurement computational-basis state and is used as the control of CNOT operations that initialize the reset state register for the subsequent interaction, such that $\ket{s_{t+1}}=\ket{s'_t}$. If a next-state control qubit is in $\ket{0}$, the designated state qubit remains unchanged, whereas a control value of $\ket{1}$ causes the target state qubit to be flipped according to the register bit ordering used in the implementation. After state initialization, the next-state register is reset for reuse. This procedure maintains the sampled state transition between successive interaction steps without requiring an additional state register.

\begin{algorithm}[ht]
\caption{Dynamic QMDP Execution for a $T$-Step Interaction with Qubit Reuse}
\label{alg:dynamic_qmdp}
\begin{algorithmic}[1]
\small
\REQUIRE Horizon $T$
\STATE \textbf{Initialization:}
\STATE Prepare \texttt{qState} in a superposition over environment states.
\STATE Initialize \texttt{qReturn}$\leftarrow\ket{0}$.
\FOR{$t=0$ to $T-2$}
    \STATE Apply Hadamard gate(s) to \texttt{qAction} to generate an action superposition.
    \STATE Apply the transition operator $U_P$ to prepare the next-state register \texttt{qNextState}.
    \STATE Apply the reward operation to encode the corresponding reward in \texttt{qReward}.
    \STATE Transfer the reward information from \texttt{qReward} to the dedicated ancillary reward register \texttt{qAnci[$t$]}.
    \STATE Measure \texttt{qState}, \texttt{qAction}, \texttt{qNextState}, and \texttt{qReward}.
    \STATE Store the measurement outcomes in the designated bits of \texttt{cReg}.
    \STATE Reset \texttt{qState}, \texttt{qAction}, and \texttt{qReward}.
    \STATE Initialize \texttt{qState} from the post-measurement \texttt{qNextState} using CNOT operations.
    \STATE Reset \texttt{qNextState}.
\ENDFOR
\STATE \textbf{Final interaction} ($t=T-1$):
    \STATE Apply Hadamard gate(s) to \texttt{qAction}.
    \STATE Apply the transition operator $U_P$ to prepare the final \texttt{qNextState}.
    \STATE Apply the reward operation to encode the final reward in \texttt{qReward}.
\STATE Construct \texttt{qReturn} using the retained ancillary rewards and the final-step reward.
\STATE Apply Grover-based trajectory search.
\STATE Measure the final interaction registers and \texttt{qReturn}.
\STATE Combine the intermediate measurement records with the final measurement results to reconstruct the complete trajectory.
\end{algorithmic}
\end{algorithm}

The temporal evolution proceeds as follows. At the initial time step ${t=0}$, the state register is prepared in a superposition over initial environment states, and the action register is initialized in a uniform superposition, enabling parallel evaluation of the state–action pairs. 
The transition operator $U_P$ is then applied to the state and action registers to prepare the corresponding next-state register $\ket{s'_0}$, followed by the reward operation to produce $\ket{r_0}$. Before the interaction registers are measured, the reward information is transferred to a dedicated ancillary quantum register using CNOT operations. A mid-circuit measurement is subsequently performed on the state, action, next-state, and reward registers, yielding a sampled tuple $(s_0, a_0, s'_0, r_0)$. These measurement outcomes are stored in the classical memory for trajectory reconstruction. 
The measurement projects the interaction registers onto the sampled outcome. Consequently, the retained ancillary register contains the reward value associated with that sampled interaction rather than a coherent superposition over all preceding trajectory branches. Therefore, the dynamic execution does not preserve a coherent superposition of the complete trajectories across the full interaction horizon.

Immediately after measurement, the state, action, and reward registers are reset for reuse. The next-state register remains in its post-measurement computational-basis state and is used to initialize the reset state register through CNOT operations. The next-state register is then reset, and the action register is reinitialized in a uniform superposition for the subsequent interaction. This reset–propagate–reinitialize sequence maintains the sequential continuity of the interaction while reusing the same interaction-register workspace.

At time step ${t=1}$, the agent interacts with the updated state $\ket{s_1}$, the action register is prepared in superposition, and the environment generates $\ket{s'_1}$ and reward $\ket{r_1}$. As in the initial interaction, the reward information is first transferred to its dedicated ancillary register, after which the interaction registers are measured and the resulting tuple $(s_1, a_1, s'_1, r_1)$ is stored classically. This procedure is repeated for all intermediate steps \(t=0,\ldots,T-2\). At the final interaction step \(t=T-1\), the transition and reward operations are applied without an intermediate measurement, so that the final-step registers remain available for the return calculation and the subsequent search. The retained reward registers from the preceding interaction steps, together with the final-step reward register, are used to construct the quantum return register. The retained reward registers, the return register, and the final-interaction registers (\texttt{qState}, \texttt{qAction}, and \texttt{qReward}) are then used in the Grover-based trajectory search described in Section~\ref{method:return_grover}.

Thus, a single execution provides the intermediate interaction outcomes through classical measurement records, while the final-step registers and return register are measured after return evaluation and, when applied, Grover-based trajectory search. The complete interaction trajectory is reconstructed by combining the intermediate and final interaction outcomes:
\(\begin{multlined}
\tau^{(T)} = (s_0,a_0,s'_0,r_0,\;
              s_1,a_1,s'_1,r_1,\ldots,
              s_{T-1},a_{T-1},s'_{T-1},r_{T-1}).
\end{multlined}
\)
Repeating the circuit over multiple shots yields an ensemble of reconstructed trajectories. The statistical distribution of these trajectories is determined by the quantum state preparation, action preparation, and transition amplitudes within each interaction step. Therefore, the resulting measurement outcomes provide an empirical trajectory distribution over repeated circuit executions.

\subsection{Quantum Return Calculation and Trajectory Search}\label{method:return_grover}
In classical RL, the primary objective is to find the optimal policy, which dictates the best action in each state so as to maximize the long-term return. In quantum RL (QRL), quantum algorithms are applied to efficiently search for desired trajectories and to identify the optimal action for each state. In this study, Grover’s search~\cite{groveralgorithm} is utilized to identify the optimal quantum trajectories that yield the maximum return over a finite interaction horizon.
Within the proposed QRL framework, the dynamic QMDP generates state-action-reward trajectories according to the encoded transition dynamics, while the quantum return register provides a trajectory-level performance measure.  Grover’s algorithm uses the retained reward registers and the computed return, together with the state, action, and reward registers of the final interaction step, to search for the optimal quantum trajectories. These trajectories reveal the state-action sequences associated with maximum return, from which the optimal action for each state can be inferred. The following subsections describe the quantum return calculation and the Grover-based quantum trajectory search in detail.

\subsubsection{Quantum Return Calculation}
In classical RL, the reward is the key signal that guides the agent's behavior. It is a scalar value returned by the environment as feedback for the action taken in each state. The agent aims to maximize the cumulative reward over time, and the total reward accumulated over a trajectory of interactions is termed the return. The return is typically discounted by a factor $\gamma \in [0, 1]$, which weights immediate rewards more heavily than future ones.
In this work, this classical concept is mapped onto the quantum domain, where the return is encoded in a dedicated quantum register \(\ket{g}\) (\texttt{qReturn}). The register width is selected to represent the required range of return values over the finite interaction horizon, and the register is initialized to \(\ket{0_g}\) at the beginning of each circuit execution.
For a finite-horizon trajectory \(\tau^{(T)}\), the return is generally written as:
\[g\!\left(\tau^{(T)}\right)=\sum_{t=0}^{T-1}\gamma^t r_t, \]
where \(\gamma\in[0,1]\) denotes the discount factor. In the present implementation, \(\gamma=1\), giving an undiscounted return:
\[ g\!\left(\tau^{(T)}\right)=\sum_{t=0}^{T-1}r_t.\]

Because the interaction registers from the earlier steps have already been measured during the dynamic QMDP execution, the return calculation does not operate on a coherent quantum representation of the complete multi-step trajectory.  Instead, the reward information from the intermediate interaction steps is transferred to dedicated ancillary quantum registers before each intermediate measurement. These registers therefore retain the reward values associated with the sampled earlier interaction outcomes, while the final-step reward remains available in \texttt{qReward}. In the implemented three-step circuit, the rewards from the first two interactions are retained in two ancillary reward registers and combined with the final-step reward to construct \texttt{qReturn}. 

The quantum return register is constructed after the final interaction rather than updated after each interaction step. The retained rewards and the final-step reward are combined through reversible quantum arithmetic, implemented with CNOT, Toffoli and multi-controlled operations acting on the reward registers and \texttt{qReturn}. The resulting \texttt{qReturn} encodes the accumulated return associated with the reward values and is subsequently used in the Grover-based trajectory-search stage.

\subsubsection{Grover-based Quantum Trajectory Search}
\label{method:grover}
After construction of the quantum return register, the Grover-based trajectory-search stage operates on the quantum information that remains available after the dynamic interaction sequence and the return calculation. The earlier interaction registers have already been measured and reset; therefore, this stage does not operate on a coherent quantum register containing the complete multi-step trajectory. Instead, the retained reward registers, the return register, and the final-interaction registers (\texttt{qState}, \texttt{qAction}, and \texttt{qReward}) participate in the implemented search procedure.

For the finite benchmark considered in this study, the maximum achievable return is known a priori and is encoded directly in the marking condition of the phase-marking oracle. The oracle therefore marks the configurations whose return corresponds to this value, and Grover’s search amplifies the probability amplitudes of these marked configurations, enabling the identification of the optimal trajectories that achieve this return. 
Let \(\ket{x}\) represent a computational-basis configuration of the registers participating in the oracle, and let \(f(x)\) be a Boolean marking function such that \(f(x)=1\) for a marked configuration and \(f(x)=0\) otherwise. The phase-marking oracle \(U_w\) is defined as:
\begin{equation}
U_w\ket{x}=
\begin{cases}
-\ket{x}, & f(x)=1,\\
\phantom{-}\ket{x}, & f(x)=0.
\end{cases}
\label{eq:U_f}
\end{equation}

In the implemented circuit, the marking condition is realized through basis-state conditioning on the retained reward registers, the final-step reward register, and the quantum return register, followed by a multi-controlled phase inversion. No separate reversible-comparator ancilla is required because this benchmark-specific maximum return value is encoded directly in the oracle. Thus, the oracle changes the phase of the marked configurations while leaving their computational-basis values unchanged.

Following the phase-marking step, a diffusion operation is applied to amplify the probability amplitudes of the marked configurations relative to those of the unmarked configurations. In the implemented circuit, the diffusion stage consists of Hadamard, Pauli-\(X\), and multi-controlled operations acting on the quantum registers participating in the search. Importantly, this diffusion operation is not constructed by inverting the complete dynamic QMDP evolution. The preceding mid-circuit measurements and reset operations are non-unitary and therefore do not admit a unitary inverse over the complete interaction history. Accordingly, the diffusion operation acts only on the quantum registers available during the final search stage, including the state and action registers, retained reward registers, final-step reward register, and quantum return register.

Finally, the quantum registers participating in the search are measured, and the resulting measurement outcomes are combined with the interaction records stored from the preceding steps, which provide the intermediate state-action sequences, to reconstruct the corresponding complete state-action trajectories. Because the search amplifies the configurations marked by the oracle, these configurations dominate the measurement statistics over repeated circuit executions. The optimal quantum trajectories yielding the maximum return are therefore identified as the reconstructed trajectories that occur with the highest frequency.

\section{Demonstrations and Results}\label{sec:experiments_results}
The proposed QRL framework was evaluated using a four-state, two-action QMDP implemented with dynamic-circuit-based qubit reuse. The experiments assessed the execution of multi-step agent-environment interactions using a reusable interaction workspace and compared its qubit resource requirements with those of the prior static QMDP implementation~\cite{thet2024}. Ideal simulations were used to evaluate the trajectory behavior of the dynamic implementation and its agreement with the corresponding static formulation. Grover-based trajectory search was then incorporated to identify the optimal trajectories that maximize the overall return, thereby supporting the identification of the optimal action for each state. Finally, the framework was executed on quantum hardware to evaluate the feasibility of the proposed dynamic-circuit implementation under realistic noise conditions.

\subsection{Formulation of Quantum Markov Decision Process}
\begin{figure}[H]
    \centering
    \includegraphics[width = 6.85cm]{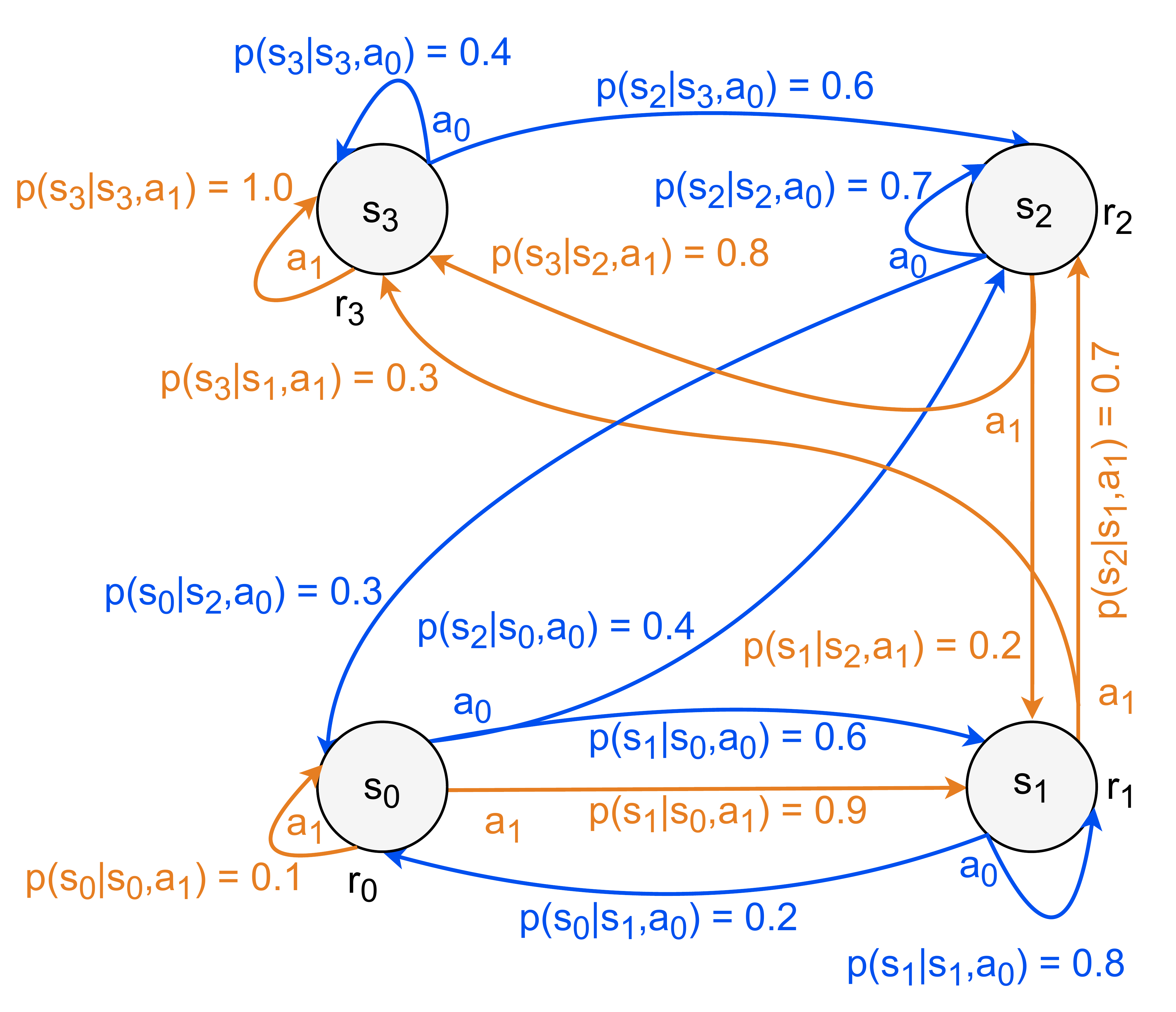}
    \vspace{-10pt}
    \caption{Graphical representation of a classical MDP with four states ($s_0$–$s_3$), two actions $(a_0, a_1)$, and rewards ($r_0$–$r_3$). Arrows denote state transitions with probabilities $P(s'|s,a)$. Transitions under action $a_0$ are shown in blue, and those under action $a_1$ are shown in orange.}
    \label{fig:state_transition}
\end{figure}

This subsection introduces the classical MDP model and its quantum encoding, which forms the basis of the subsequent dynamic-circuit implementation. Figure~\ref{fig:state_transition} illustrates the state–transition and reward structure of the classical MDP, which serves as the reference model for quantum circuit implementation.

The environment consists of four discrete states, $s_0,s_1,$ $s_2,s_3$, and two actions, $a_0$ and $a_1$. Arrows between states represent possible transitions under each action, labeled with probabilities $P(s'|s,a)$ that capture the stochastic nature of the dynamics. For example, from $s_0$, action $a_0$ leads to $s_1$ with probability 0.6 or to $s_2$ with 0.4, whereas action $a_1$ returns to $s_0$ with probability 0.1, or advances to $s_1$ with 0.9. Analogous probabilistic rules govern the other states, with $s_3$ functioning as a designated target state whose evolution depends on the chosen action and its transition probabilities. Each state is associated with a corresponding reward $(r_0, r_1, r_2, r_3)$, indicated alongside the nodes. These rewards define the state-based reward function and determine the return accumulated along a trajectory, which serves as the performance measure in the subsequent trajectory-search stage. Collectively, the state space, action space, transition probabilities, and reward function define the classical MDP encoded within the proposed dynamic-circuit QMDP framework.

\subsubsection{Quantum Encoding of the Classical MDP}
\begin{figure}[H]
    \includegraphics[width= \textwidth]{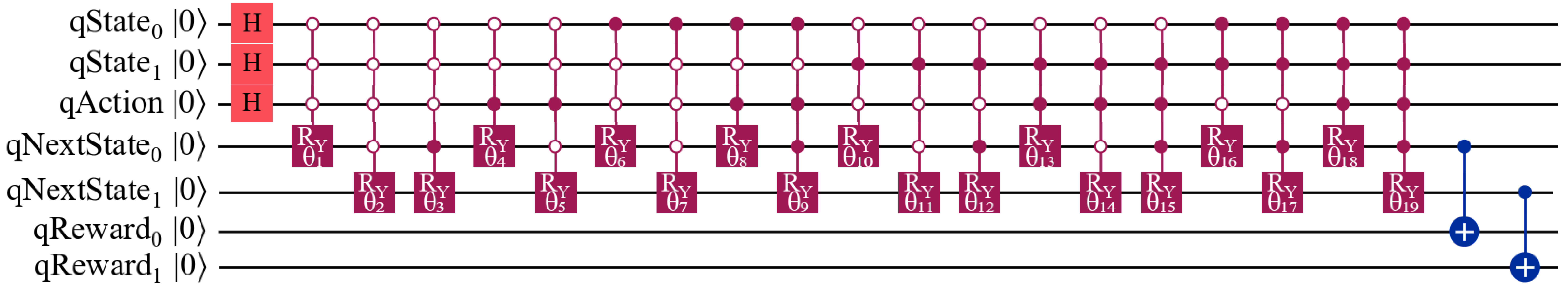}
    \vspace{-15pt}
    \caption{Quantum circuit for the benchmark MDP encoding states, actions, stochastic transitions, and rewards into quantum registers. Controlled $R_y(\theta)$ rotations encode the transition probabilities \(P(s'|s,a)\) in the next-state register, while CNOT gates encode rewards conditioned on the next states.}
    \label{fig:one_step}
\end{figure}

The classical MDP described in Figure~\ref{fig:state_transition} was implemented in the quantum domain by encoding its dynamics into quantum states. This quantum realization, as shown in Figure~\ref{fig:one_step}, preserved the structure of the classical model while using superposition to represent multiple state--action transition outcomes within a single QMDP interaction.

This implementation used four quantum registers. The state register (\texttt{qState}) consisted of two qubits that encoded the four states $(s_0,s_1,s_2,s_3)$. For the initial interaction, the register was initialized in $\ket{00}$ and transformed into a uniform superposition using Hadamard gates. The action register (\texttt{qAction}) used a single qubit, initialized in $\ket{0}$ and placed in a uniform superposition to represent both candidate actions. The next-state register (\texttt{qNextState}), composed of two qubits initialized in $\ket{00}$, was prepared using multi-controlled $R_y(\theta)$ rotations whose angles were determined by the conditional probabilities of successive branches in the next-state preparation. The rotations were conditioned on the computational-basis values of the state and action registers, with additional next-state controls introduced where required by the recursive preparation procedure. The reward register (\texttt{qReward}), also initialized in $\ket{00}$, was encoded using CNOT operations from the next-state register, implementing the benchmark reward mapping associated with the resulting state. 

This quantum representation jointly encoded the state, action, transition, and reward information of a single QMDP interaction within the corresponding quantum registers. Superposition allowed multiple state--action transition outcomes of the interaction to be represented simultaneously within the same quantum state. The resulting QMDP circuit provides the basic interaction block, which is extended to multiple time steps through dynamic-circuit-based qubit reuse in Section~\ref{result:multi_step}.

\subsection{Experimental Setup and Reproducibility}
The quantum circuits were implemented using Qiskit 2.0.0, Qiskit Aer 0.17.0, Qiskit Runtime 0.37.0, and Qiskit Experiments 0.10.0. Ideal simulations were performed using Qiskit Aer with 1,024 shots per circuit execution, and no hardware-derived noise model was applied. Hardware experiments were conducted on the 133-qubit IBM Heron-class \texttt{ibm\_torino} processor using the Qiskit Runtime Sampler with 32,768 shots per circuit execution.  All quantum registers were initialized to the computational-basis state \(\lvert 0\rangle\) before circuit execution. A delay of 2,000 ns was inserted between the mid-circuit measurement and the subsequent reset and qubit reuse during hardware execution.

The circuits were transpiled to the target IBM backend using Qiskit's preset pass manager with an optimization level of 3. No explicit layout method, routing method, or transpiler seed was specified; therefore, these choices were determined by the preset pass-manager configuration associated with the employed Qiskit version and backend. The transpiled circuits used the native \(\{\mathrm{rz},\mathrm{sx},\mathrm{x},\mathrm{cz}\}\) gate set together with measurement, reset, and delay instructions. XY4 dynamical decoupling was enabled during hardware execution.

Before transpilation, the three-step dynamic QMDP circuit without Grover-based trajectory search used 15 logical qubits, had a circuit depth of 107, a circuit size of 159 operations, and contained 106 quantum gate operations. The corresponding circuit with Grover-based trajectory search also used 15 logical qubits, with a pre-transpilation depth of 131, a circuit size of 209 operations, and 156 quantum gate operations. Both circuits contained 25 measurement, 14 reset, and 14 delay instructions before transpilation.

For the dynamic QMDP without Grover-based trajectory search, the transpiled circuit had a depth of \(8{,}828\) and a circuit size of \(14{,}527\). It contained \(7{,}180\) \(\mathrm{sx}\), \(3{,}302\) \(\mathrm{rz}\), \(172\) \(\mathrm{x}\), and \(3{,}820\) \(\mathrm{cz}\) gates, corresponding to \(10{,}654\) one-qubit native gates and \(3{,}820\) two-qubit gates. The circuit additionally contained 25 measurement, 14 reset, and 14 delay instructions. The corresponding archived IBM Runtime job reported a quantum usage time of 36 seconds.
For the dynamic QMDP with Grover-based trajectory search, the transpiled circuit had a depth of \(10{,}566\) and a circuit size of \(17{,}710\). It contained \(8{,}898\) \(\mathrm{sx}\), \(3{,}865\) \(\mathrm{rz}\), \(191\) \(\mathrm{x}\), and \(4{,}703\) \(\mathrm{cz}\) gates, corresponding to \(12{,}954\) one-qubit native gates and \(4{,}703\) two-qubit gates. The circuit additionally contained 25 measurement, 14 reset, and 14 delay instructions.  The corresponding archived IBM Runtime job reported a quantum usage time of 51 seconds.

The no-Grover and Grover circuits were mapped to backend-specific physical qubits during optimization-level-3 transpilation. For the no-Grover circuit, 15 logical qubits were initially placed on physical qubits \(43\text{--}50\), \(55\), and \(64\text{--}69\), with physical qubit \(56\) additionally used during routing. The effective CZ connectivity traversed by this circuit comprised the edges \((43,44)\), \((44,45)\), \((45,46)\), \((46,47)\), \((46,55)\), \((47,48)\), \((48,49)\), \((49,50)\), \((50,56)\), \((55,65)\), \((56,69)\), \((64,65)\), \((65,66)\), \((66,67)\), \((67,68)\), and \((68,69)\).
For the Grover circuit, 15 logical qubits were initially placed on physical qubits \(42\text{--}45\), \(54\), \(58\text{--}65\), \(72\), and \(73\), with routing through neighboring physical qubits as required by the backend connectivity. The effective CZ connectivity traversed by the circuit comprised the edges \((42,43)\), \((42,54)\), \((43,44)\), \((44,45)\), \((45,46)\), \((46,55)\), \((54,61)\), \((55,65)\), \((58,59)\), \((59,60)\), \((59,72)\), \((60,61)\), \((61,62)\), \((62,63)\), \((63,64)\), \((63,73)\), and \((64,65)\).

\subsection{Demonstration of Multiple Agent--Environment Interactions with Dynamic-Circuit Capability}\label{result:multi_step}
This subsection demonstrates the extension of the QMDP interaction to multiple time steps using dynamic-circuit-based qubit reuse. The implemented circuit performed three sequential agent--environment interactions while reusing the same state, action, next-state, and reward registers across the interaction steps. Figure~\ref{fig:two_state_block} illustrates the reusable QMDP interaction workspace across the three-step sequence. Mid-circuit measurement, reset, and next-state propagation enabled reuse of the interaction registers while maintaining the state transition between successive steps.

\begin{figure}[H]
    \includegraphics[width=\textwidth]{figures/Fig_3_Two_state_MDP_blocks_Dynamic.pdf}
    \vspace{-15pt}
    \caption{Reusable QMDP interaction workspace across three agent--environment interaction steps. Each colored block represents an interaction step \((t=0,1,2)\). Mid-circuit measurement and reset enable reuse of the \texttt{qState}, \texttt{qAction}, \texttt{qNextState}, and \texttt{qReward} registers across successive steps. After each intermediate measurement, the next-state register remains in its post-measurement computational-basis state and is used through CNOT operations to initialize the state register for the subsequent interaction. This qubit-reuse procedure avoids replication of the interaction registers across time steps and maintains a fixed reusable interaction-workspace width.}
    \label{fig:two_state_block}
\end{figure}

Following the interaction at \(t=0\), the reward information required for subsequent return evaluation was transferred from \texttt{qReward} to a dedicated ancillary reward register before the interaction registers were measured. The state, action, next-state, and reward registers were then measured, and the outcomes were stored in the designated bits of the classical register \texttt{cReg}. The state, action, and reward registers were subsequently reset. The next-state register remains in its post-measurement computational-basis state and was used as the control of CNOT operations to initialize the state register for the subsequent interaction, after which the next-state register was reset for reuse. The action register was then reinitialized in a uniform superposition before the next transition and reward operations were applied. The same intermediate-step procedure, including retention of the reward information before measurement, was performed after \(t=1\).

Figure~\ref{fig:two_state_block} deliberately focuses on the reusable QMDP interaction workspace represented by \texttt{qState}, \texttt{qAction}, \texttt{qNextState}, and \texttt{qReward}. In the complete implementation, the rewards from the first two interaction steps were retained in dedicated ancillary registers before the corresponding mid-circuit measurements. At the final interaction step \(t=2\), the transition and reward operations were applied without an intermediate measurement preceding return evaluation. The rewards retained from the first two interactions together with the final-step reward were then used to construct the quantum return register, followed by Grover-based trajectory search and final measurement. Thus, the three-step interaction sequence was executed within a single dynamic circuit while the same interaction workspace was reused across successive steps.

The reusable interaction workspace requires seven qubits: two for \texttt{qState}, one for \texttt{qAction}, two for \texttt{qNextState}, and two for \texttt{qReward}. In the corresponding static construction~\cite{thet2024}, separate interaction registers are allocated at each time step, requiring \(7T\) interaction-workspace qubits. For \(T=3\), this corresponds to 21 interaction-workspace qubits in the static construction compared with seven qubits in the dynamic formulation, representing a 66\% reduction in the interaction-workspace width. This reduction results from reusing the same interaction registers through mid-circuit measurement, reset, and next-state propagation rather than allocating separate registers for every time step. The retained reward registers and quantum return register are additional resources and are accounted for separately in the total quantum width. Consequently, the interaction workspace remains constant with the interaction horizon, whereas the total quantum-resource requirement additionally includes the horizon-dependent retained reward storage. A static three-step QMDP implementation, which illustrates the repeated interaction-register allocation required by static circuit unrolling, is provided in Appendix~\ref{appendixA} (Figure~\ref{fig:two_state_block_static}).

Following the three interaction steps, the trajectory return was evaluated using the retained rewards from the first two interactions together with the final-step reward. The implementation used a discount factor of \(\gamma=1\), so all three rewards contributed equally to the return. As described in the methods, these reward values were combined in \texttt{qReturn} after the final interaction through reversible quantum arithmetic implemented using CNOT, Toffoli, and multi-controlled operations. The return register was constructed after completion of the interaction sequence rather than updated incrementally at each time step.

\begin{figure}[ht]
    \includegraphics[width=\textwidth]{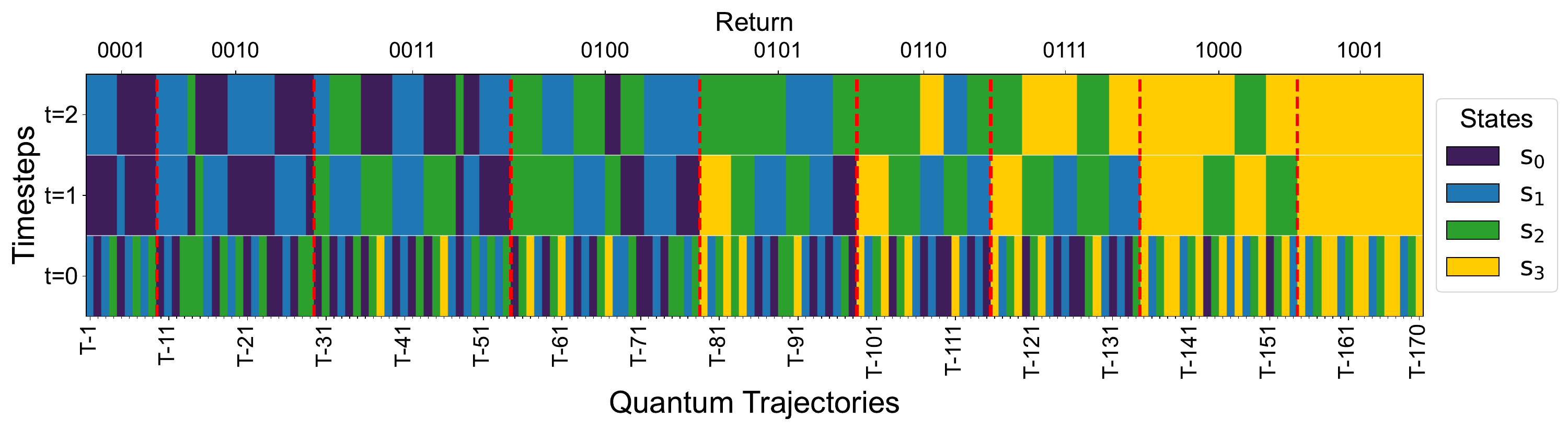}
    \vspace{-20pt}
    \caption{
    Visualization of state-visitation patterns across three time steps for quantum trajectories generated by the dynamic QMDP circuit in an ideal noise-free simulation. Columns represent individual trajectories, rows denote interaction steps, and colors indicate visited states. Trajectories are grouped by their total returns, with red dashed lines indicating group boundaries and binary strings above the panel denoting the corresponding return values.}
    \label{fig:simulation}
\end{figure}

To assess the functional behavior of the dynamic QMDP formulation, the trajectory outcomes generated by the complete three-step interaction sequence were analyzed using an ideal, noise-free Qiskit Aer simulation. Figure~\ref{fig:simulation} shows the state-visitation patterns of the sampled trajectories. Each column represents a trajectory, and each row corresponds to an interaction step \((t=0,1,2)\). The colors indicate the environment states \(s_0\)--\(s_3\), while the trajectories are grouped according to their total return. The diversity of state-visitation patterns reflects the stochastic transition dynamics and candidate actions sampled from the uniform action superposition. Trajectories with higher returns more frequently include visits to \(s_3\), which is associated with the highest reward in the benchmark reward structure.

To compare the static and dynamic QMDP implementations, the static QMDP results presented in Table~\ref{table:classical_trajectories} were used as the reference. The 170 trajectories generated by the static circuit were enumerated as state--action--next-state--reward sequences over the three interaction steps and grouped according to their accumulated returns. For the dynamic implementation, the state sequences were extracted from the state-visitation patterns shown in Fig.~\ref{fig:simulation}. The corresponding actions and rewards were then identified using the benchmark MDP transition dynamics and reward structure, respectively, allowing the complete trajectories and their accumulated returns to be reconstructed. These reconstructed trajectories were compared with the static reference trajectories in terms of their interaction-step sequences, completed three-step trajectories, and accumulated returns. The comparison showed that both implementations generated the same set of 170 admissible trajectories with corresponding return assignments, indicating that dynamic-circuit qubit reuse preserved the valid trajectory structure and sequential transition process of the benchmark. The probability distributions of these trajectories were examined separately in Section~\ref{staticvsdynamic}, where their statistical agreement was evaluated using trajectory probabilities, expected returns, and total variation distance.

\subsubsection{Comparative Evaluation: Static and Dynamic QMDP Implementations}\label{staticvsdynamic}
This subsection compares the proposed dynamic-circuit implementation with the static unrolled formulation in Ref.~\cite{thet2024} under the same benchmark definition. Both implementations use the same state and action spaces, transition probabilities, reward function, and three-step interaction horizon. The static baseline circuit is reproduced in Appendix~\ref{appendixA} (Figure~\ref{fig:two_state_block_static}), together with the corresponding trajectory enumeration in Table~\ref{table:classical_trajectories}.

\begin{table}[H]
\caption{Comparison between static and dynamic QMDP implementations
under identical MDP settings ($T=3$).}
\label{table:comparison}
\footnotesize
\renewcommand{\arraystretch}{1}
\setlength{\tabcolsep}{5pt}

\begin{tabularx}{\textwidth}{
    @{}
    >{\raggedright\arraybackslash}p{0.27\fulllength}
    >{\raggedright\arraybackslash}p{0.22\fulllength}
    >{\raggedright\arraybackslash}X
    @{}
}
\toprule
\centering \textbf{Aspect} &
\centering \textbf{Static QMDP (Ref.~\cite{thet2024})} &
\multicolumn{1}{c}{\textbf{Dynamic QMDP (This study)}}
\\
\midrule
\textbf{Interaction horizon} ($T$)
& 3
& 3 \\
\textbf{Interaction-register qubits}
& $21\;(7\times T)$
& 7 (reused across interaction steps) \\
\textbf{Interaction-workspace scaling with $T$}
& Linear in $T$
& Constant: $2n_s+n_a+n_r$ \\
\textbf{Additional quantum registers}
& A 4-qubit return register
& Two 2-qubit retained reward registers and a 4-qubit return register \\
\textbf{Total quantum width}
& 25 qubits
& 15 qubits \\
\textbf{Quantum trajectory set}
& Table~\ref{table:classical_trajectories}
& Same trajectory set in ideal simulation \\
\textbf{Quantum hardware execution}
& Not demonstrated
& Executed on IBM Heron-class hardware \\
\textbf{Circuit paradigm}
& Static
& Dynamic with mid-circuit measurement, reset, and qubit reuse \\
\bottomrule
\end{tabularx}
\end{table}

Table~\ref{table:comparison} summarizes the principal structural and resource differences between the two implementations. For \(T=3\), the static QMDP requires 21 interaction-workspace qubits because a separate seven-qubit interaction block is allocated at each time step. In the dynamic formulation, the same seven interaction qubits are reused through mid-circuit measurement, reset, and next-state propagation. This seven-qubit count refers specifically to the reusable interaction workspace. The complete dynamic implementation additionally includes two 2-qubit retained reward registers and a 4-qubit quantum return register, giving a total logical width of 15 qubits for the three-step benchmark. Comparison of the ideal dynamic-circuit results shown in Figure~\ref{fig:simulation} with the static QMDP baseline presented in Table~\ref{table:classical_trajectories} confirmed that both implementations generated the same set of 170 admissible three-step trajectories with identical return values.

To provide a quantitative comparison under a common benchmark setting, the static and dynamic QMDP implementations were evaluated for the same three-step MDP. The ideal quantum simulations used 1,024 shots per circuit execution. For statistical evaluation, 30 independent simulator seeds were used, and the results are reported as mean \(\pm\) standard deviation together with 95\% confidence intervals where applicable. 
For a trajectory \(\tau\), the empirical probability was calculated as \(P(\tau)=C(\tau)/1024\), where \(C(\tau)\) denotes the number of occurrences of \(\tau\) among the 1,024 shots. The difference between the static and dynamic distributions was quantified using the total variation distance (TVD) (Equation (\ref{eq:tvd})),  where \(P_{\mathrm{static}}(\tau)\) and \(P_{\mathrm{dynamic}}(\tau)\) denote the empirical probabilities obtained from the two implementations. 
\begin{equation}
   D_{\mathrm{TV}}(P_{\mathrm{static}},P_{\mathrm{dynamic}})
=
\frac{1}{2}\sum_{\tau}
\left|P_{\mathrm{static}}(\tau)-P_{\mathrm{dynamic}}(\tau)\right|. 
\label{eq:tvd}
\end{equation}

Across the 30 simulator seeds, the mean TVD between the static and dynamic trajectory distributions was \(0.2219\pm0.0143\), indicating measurable differences in the empirical trajectory probabilities despite both implementations generating the same set of admissible trajectories and corresponding return values. 
The corresponding mean expected returns were \(5.9765\pm0.0709\) for the static QMDP and \(5.9802\pm0.0675\) for the dynamic QMDP, with 95\% confidence intervals of \([5.9501,6.0030]\) and \([5.9550,6.0054]\), respectively. The closely matched expected returns provide an additional quantitative comparison between the two execution models. The two finite-shot implementations should therefore not be interpreted as producing identical probability distributions.

Overall, the comparison shows that the proposed dynamic QMDP preserves the admissible trajectory set and closely matches the expected return of the static baseline while reducing the interaction-workspace requirement from 21 to seven qubits and the total quantum width from 25 to 15 qubits. By reusing the interaction registers across sequential steps, the formulation reduces the physical-qubit overhead associated with static circuit unrolling while preserving the sequential structure of the benchmark QMDP. This provides a more resource-efficient approach to finite-horizon multi-step QRL and may improve the feasibility of such implementations on NISQ hardware.

\subsubsection{Implementation of Multiple Interactions on Quantum Hardware}
To evaluate the execution of the proposed dynamic-circuit QMDP on quantum hardware, a complete three-step interaction circuit was implemented on the 133-qubit IBM Heron-class \texttt{ibm\_torino} processor. The hardware experiment was used to assess whether the sequential dynamic-circuit procedure, including mid-circuit measurement, reset, next-state propagation, and qubit reuse, could be executed on a superconducting quantum processor under realistic device noise. The hardware execution required additional timing considerations that are absent from the ideal simulation. In particular, repeated measurement, reset, and reuse of the interaction registers introduced timing constraints between successive interaction steps. Accordingly, a delay of 2,000 ns was inserted between the mid-circuit measurement and subsequent reset and qubit reuse during hardware execution. This delay provided an explicit separation between the measurement and reuse operations within the dynamic-circuit sequence.

\begin{figure}[H]
    \includegraphics[width=\textwidth]{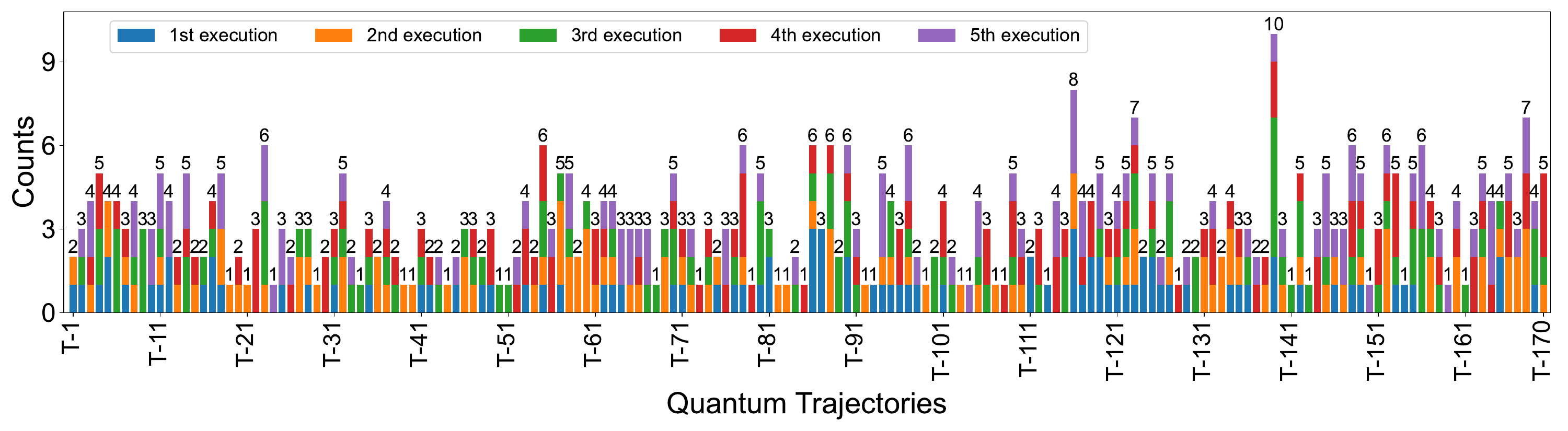}
    \vspace{-20pt}
    \caption{Filtered trajectory distributions obtained from five repeated hardware experiments of the three-step dynamic QMDP on the 133-qubit IBM Heron-class \texttt{ibm\_torino} processor. Each bar corresponds to one of the 170 reference trajectories obtained from the ideal three-step QMDP simulation, with stacked colors indicating the retained counts from the first through fifth hardware experiments. The displayed counts include only hardware outcomes matching this reference trajectory set and therefore do not represent the complete raw hardware measurement results.}
    \label{fig:ibm}
\end{figure}

For the hardware evaluation of the multi-step dynamic QMDP, the experiment was repeated five times, with 32,768 shots per circuit execution. The resulting hardware measurements contained both outcomes that corresponded to the trajectories observed in the ideal three-step QMDP simulation and additional outcomes arising under hardware execution. To analyze the hardware trajectories, the measurement results were filtered using 170 valid trajectories obtained from the ideal three-step simulation, including the associated return values, as the reference set. These trajectories correspond to \(T\)-1 through \(T\)-170 in Figure~\ref{fig:simulation}. Only hardware outcomes matching this reference trajectory set were retained and displayed in Figure~\ref{fig:ibm}.

Figure~\ref{fig:ibm} presents the retained trajectory counts across the five hardware experiments. Each bar corresponds to one valid trajectory, while the stacked colors show the observed counts from the first through the fifth experiments. Across the five repeated experiments, every trajectory in the valid reference set was observed at least once, demonstrating that the expected trajectory configurations could be observed on hardware despite noise. However, the coverage of the valid trajectory set does not imply agreement with the ideal sampling distribution. The observed counts varied substantially across the trajectories and repeated experiments, reflecting the effect of hardware noise on the measured trajectory probabilities.

Accordingly, the hardware experiment is interpreted as a feasibility demonstration of the proposed multi-step dynamic-circuit execution model rather than as evidence of close agreement with the ideal trajectory distribution. The experiment demonstrates that mid-circuit measurement, reset, next-state propagation, and interaction-register reuse can be executed within the proposed three-step QMDP workflow on the target processor. 

\subsection{Experimental Results on Grover-based Quantum Trajectory Search}\label{sec:grover}
This subsection evaluates the Grover-based trajectory search  using the three-step benchmark. The objective is to identify optimal quantum trajectories that yield the maximum return over the finite interaction horizon under a specified scenario. Specifically, the search considers trajectories in which the agent starts from state \(s_0\) (‘00’) and reaches state \(s_3\) (‘11’) while achieving the maximum cumulative return. The corresponding optimal state--action sequences are subsequently examined to identify the actions associated with the maximum-return trajectories. 

The ideal three-step QMDP simulation yielded a set of 170 trajectories spanning the benchmark state and action configurations. For this analysis, attention was restricted to trajectories beginning at \(s_0\) (‘00’) and ending at \(s_3\) (‘11’), as summarized in Table~\ref{table:trajectories}. Among the 11 trajectories satisfying this condition, two achieved the maximum return \(g^\star=8\) (binary ‘1000’): T-151 (‘1000111111111111101010000’) and T-143 (‘1000111101111111101010000’). These two trajectories represent the maximum-return candidates associated with the return condition considered in the Grover-based search.
For this finite benchmark, the maximum return \(g^\star= \text{`1000'}\) was known a priori from the defined reward structure and was directly encoded in the oracle. One Grover iteration was applied, consisting of one phase-marking oracle invocation followed by one diffusion operation. At the logical-circuit level, the oracle consisted of six Pauli-\(X\) gates, two Hadamard gates, and one eight-controlled \(X\) operation. The diffusion operation comprised eight Pauli-\(X\) gates, 24 Hadamard gates, five three-controlled \(X\) operations, and four five-controlled \(X\) operations. 

\begin{table*}[!ht]
\centering
\begin{minipage}[t]{0.58\textwidth}
\vspace{0pt}
\captionof{table}{Trajectories originating at $s_0$ and ending at $s_3$ over three interaction steps,
grouped by return value.}
\label{table:trajectories}
\centering
\tiny
\setlength{\tabcolsep}{2pt}
\renewcommand{\arraystretch}{1.05}
\resizebox{\linewidth}{!}{%
\begin{tabular}{c c c c c}
\toprule
\textbf{Trajectory no.} &
\textbf{Quantum trajectory} &
\textbf{$t=2$} &
\textbf{$t=1$} &
\textbf{$t=0$} \\
\midrule
\rowcolor{groupgray}
\multicolumn{5}{l}{\textbf{Return: 1000}}\\
T-151 & 1000111111111111101010000 & 1111111 & 1111110 & 1010000 \\
T-143 & 1000111101111111101010000 & 1111011 & 1111110 & 1010000 \\
\addlinespace[1pt]
\rowcolor{groupgray}
\multicolumn{5}{l}{\textbf{Return: 0111}}\\
T-133 & 0111111111111111010101100 & 1111111 & 1111101 & 0101100 \\
T-131 & 0111111111111111010101000 & 1111111 & 1111101 & 0101000 \\
T-127 & 0111111111010100101010000 & 1111110 & 1010010 & 1010000 \\
T-126 & 0111111101111111010101100 & 1111011 & 1111101 & 0101100 \\
T-124 & 0111111101111111010101000 & 1111011 & 1111101 & 0101000 \\
\addlinespace[1pt]
\rowcolor{groupgray}
\multicolumn{5}{l}{\textbf{Return: 0110}}\\
T-115 & 0110111111010101010101100 & 1111110 & 1010101 & 0101100 \\
T-113 & 0110111111010101010101000 & 1111110 & 1010101 & 0101000 \\
\addlinespace[1pt]
\rowcolor{groupgray}
\multicolumn{5}{l}{\textbf{Return: 0101}}\\
T-95 & 0101111110101010010101100 & 1111101 & 0101001 & 0101100 \\
T-93 & 0101111110101010010101000 & 1111101 & 0101001 & 0101000 \\
\bottomrule
\end{tabular}%
}
\end{minipage}%
\hfill
\begin{minipage}[t]{0.40\textwidth}
\vspace{0pt}
\captionof{table}{Qubit representation of quantum samples
grouped by initial state.}
\label{table:qsampledistribution}
\centering
\scriptsize
\setlength{\tabcolsep}{3pt}
\renewcommand{\arraystretch}{1.03}
\begin{tabular}{c c c c}
\toprule
\textbf{Quantum sample} &
\textbf{Reward} &
\textbf{Next state} &
\textbf{Action} \\
\midrule
\rowcolor{groupgray}
\multicolumn{4}{l}{\textbf{Initial state: 00}}\\
0000100 & 00 & 00 & 1 \\
1010000 & 10 & 10 & 0 \\
0101000 & 01 & 01 & 0 \\
0101100 & 01 & 01 & 1 \\
\addlinespace[1pt]
\rowcolor{groupgray}
\multicolumn{4}{l}{\textbf{Initial state: 01}}\\
0000001 & 00 & 00 & 0 \\
1010101 & 10 & 10 & 1 \\
0101001 & 01 & 01 & 0 \\
1111101 & 11 & 11 & 1 \\
\addlinespace[1pt]
\rowcolor{groupgray}
\multicolumn{4}{l}{\textbf{Initial state: 10}}\\
0000010 & 00 & 00 & 0 \\
1010010 & 10 & 10 & 0 \\
0101110 & 01 & 01 & 1 \\
1111110 & 11 & 11 & 1 \\
\addlinespace[1pt]
\rowcolor{groupgray}
\multicolumn{4}{l}{\textbf{Initial state: 11}}\\
1010011 & 10 & 10 & 0 \\
1111011 & 11 & 11 & 0 \\
1111111 & 11 & 11 & 1 \\
\bottomrule
\end{tabular}
\end{minipage}
\end{table*}

\begin{figure}[H]
    \includegraphics[width=\textwidth]{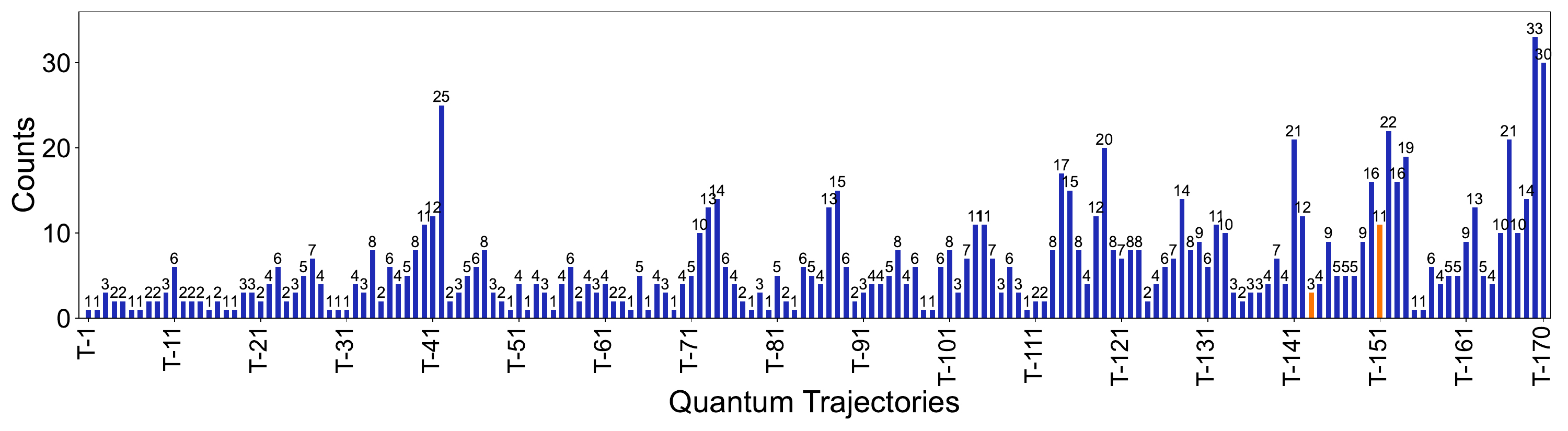}
    \vspace{-20pt}
   \caption{Trajectory distribution obtained from the three-step dynamic QMDP before applying Grover-based trajectory search in the noise-free simulation. The maximum-return trajectories T-151 and T-143, which are targeted for amplification by Grover's search, are highlighted in orange.}
    \label{fig:before_grover}
\end{figure}
\begin{figure}[H]
    \includegraphics[width=\textwidth]{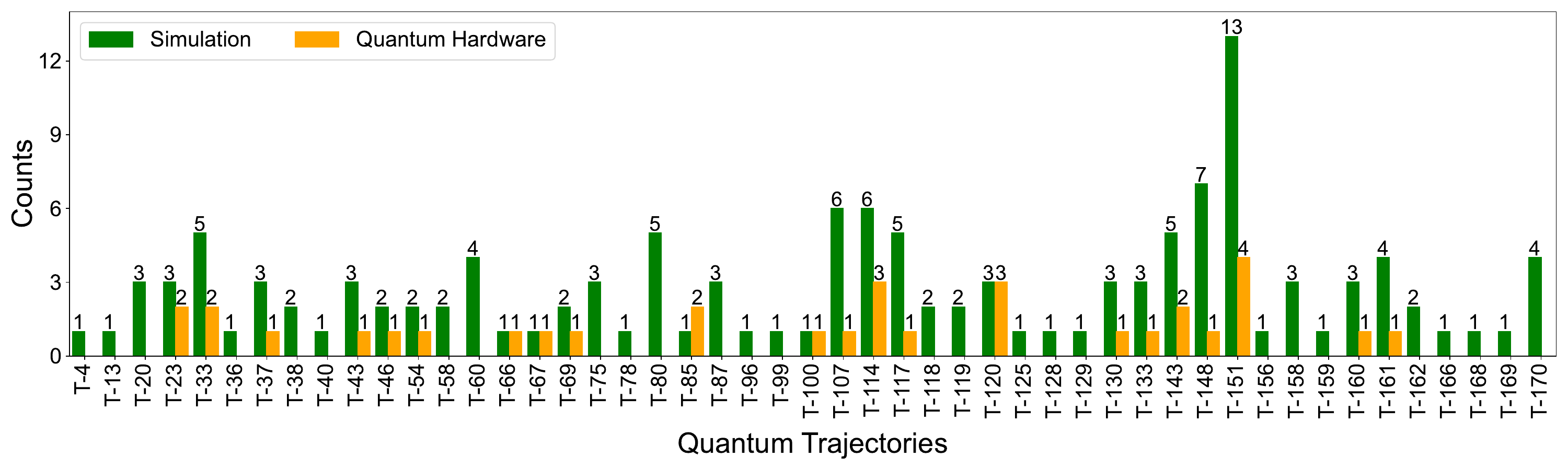}
    \vspace{-20pt}
   \caption{Quantum trajectory distribution after Grover's search. The horizontal and vertical axes represent trajectory identifiers and measurement counts, respectively. Noise-free simulation and IBM \texttt{ibm\_torino} hardware results are shown in green and yellow, respectively. T-151 and T-143 are the maximum-return trajectories targeted by the Grover search. Only outcomes matching the 170 reference trajectories from the ideal three-step dynamic QMDP simulation are displayed; thus, the counts do not represent the complete raw outputs.}
    \label{fig:grover}
\end{figure}

Figure~\ref{fig:before_grover} presents the trajectory distribution obtained from the dynamic QMDP before applying Grover-based trajectory search, whereas Figure~\ref{fig:grover} presents the corresponding distribution after Grover iteration, including both noise-free simulation and hardware execution. The horizontal axes represent the trajectory identifiers, and the vertical axes indicate the corresponding measurement counts. For the results after Grover's search, the measurement outcomes from both the noise-free simulation and hardware execution were filtered using the 170 trajectories obtained from the ideal three-step dynamic QMDP simulation as the reference set. Only outcomes matching these reference trajectories were retained for the trajectory-level comparison. Consequently, the displayed counts in Figure~\ref{fig:grover} do not represent the complete raw measurement outputs.

The comparison between the two distributions demonstrates the effect of Grover-based amplitude amplification on the identification of maximum-return trajectories. Before Grover's search, as shown in Figure~\ref{fig:before_grover}, the two maximum-return trajectories, T-151 and T-143, were not dominant in the overall trajectory distribution, as several other trajectories exhibited higher measurement frequencies. Following the application of Grover iteration, however, the measurement distribution changed substantially, as shown in Figure~\ref{fig:grover}. In the noise-free simulation, T-151 became the most frequently observed trajectory among all retained outcomes, while the measurement frequencies of other trajectories were suppressed. Although T-143 remained among the observed outcomes, its measurement frequency was comparable to those of several other trajectories and did not exhibit the same dominance as T-151. This redistribution of measurement outcomes reflects the amplitude amplification of the marked maximum-return configurations, particularly T-151, thereby increasing their measurement prominence relative to the other trajectories. The hardware results are examined separately in Section~\ref{grover_hardware}.

To further interpret the identified optimal trajectories, T-151 is examined in detail using its corresponding interaction-level representation in Table~\ref{table:trajectories}. At \(t=0\), the trajectory begins at \(s_0\) (‘00’), where action \(a_0\) (‘0’) is selected. The transition leads to \(s_2\) (‘10’), yielding reward \(r_2\) (‘10’), as represented by the seven-bit sub-string ‘1010000’. At \(t=1\), action \(a_1\) (‘1’) is selected from \(s_2\), leading to \(s_3\) (‘11’) and yielding reward \(r_3\) (‘11’) , represented by ‘1111110’. At \(t=2\), action \(a_1\) (‘1’) is selected again at \(s_3\), resulting in the agent remaining in \(s_3\) and receiving reward \(r_3\) (‘11’), represented by ‘1111111’. The accumulated return of this trajectory is \(g^\star= 8 (`1000')\). The corresponding qubit representation of each interaction step, decomposed into its constituent quantum samples, is presented in Table~\ref{table:qsampledistribution}. Each seven-bit quantum sample encodes, from least to most significant bits, the state (two qubits), action (one qubit), next state (two qubits), and reward (two qubits).

Based on the identified optimal trajectory T-151, the corresponding optimal action sequence under the considered scenario is \(a_0\) at \(s_0\), followed by \(a_1\) at \(s_2\) and \(a_1\) at \(s_3\). The results demonstrate that Grover-based amplitude amplification enhances the measurement prominence of the marked maximum-return trajectories, with T-151 emerging as the dominant trajectory in the noise-free simulation after one Grover iteration. The same optimal trajectory is also present in the static QMDP baseline, as presented in Appendix~\ref{appendixA}, providing a consistency check between the static and dynamic formulations.

\subsubsection{Quantum-Hardware Evaluation of Grover-based Trajectory Search}\label{grover_hardware}
To further evaluate the proposed QRL framework, the dynamic QMDP with Grover-based trajectory search was executed on the IBM \texttt{ibm\_torino} processor. Each hardware execution used 32,768 shots, and 30 executions of the Grover-based circuit were analyzed. The trajectory counts reported for the hardware experiment correspond only to measurement outcomes matching the 170 reference trajectories obtained from the ideal dynamic QMDP simulation. The complete raw hardware measurements additionally contained a large number of noise-induced bit strings outside this reference set. 

The filtered hardware distribution is shown in Figure~\ref{fig:grover}. Across the 30 hardware executions, T-151 was observed four times and T-143 twice. The remaining retained counts were distributed among other reference trajectories, while additional raw measurement outcomes outside the reference set were excluded from the plotted distribution. The sparse occurrence of the two marked trajectories and the broad distribution of other outcomes illustrate the sensitivity of the Grover-based stage to hardware errors in the dynamic circuit. 

For the previously presented hardware implementation without Grover-based search (Figure~\ref{fig:ibm}), T-143 and T-151 were observed four times in total over five executions of 32,768 shots, corresponding to a normalized marked-trajectory probability of approximately \(0.00244\%\). After applying Grover-based search, the normalized probability of the two marked trajectories was approximately \(0.000610\%\). 
Because the two hardware data sets were obtained from different numbers of independent executions, the normalized probability of sampling the marked maximum-return trajectories was calculated using the total number of raw hardware shots as the denominator, rather than the filtered trajectory counts.
Under these experimental conditions, the hardware sampling probability of the marked trajectories did not increase after application of the Grover-based search. Accordingly, this result is interpreted as an empirical hardware observation rather than as a statistically controlled estimate of the amplification effect.

Overall, the experiment provides evidence of the feasibility of executing the proposed dynamic-circuit QMDP and Grover-based trajectory-search procedure on current NISQ hardware, but does not establish hardware-level amplification or computational advantage. Although XY4 dynamical decoupling was enabled during hardware execution, a controlled comparison with and without dynamical decoupling was not performed; consequently, its quantitative contribution to the measured results is not assessed in this study.

\section{Discussion}\label{sec:discussion}
The principal contribution of this study is a resource-efficient dynamic-circuit execution model for multi-step QMDPs. A static unrolled formulation allocates a separate interaction workspace for each time step, causing the interaction-workspace requirement to grow linearly with the horizon. The proposed formulation instead reuses the same state, action, next-state, and reward registers across sequential interactions through mid-circuit measurement, reset, re-initialization, and next-state propagation. 
For the three-step benchmark, the reusable interaction workspace was reduced from 21 qubits in the static formulation to seven qubits in the dynamic implementation, corresponding to a 66\% reduction in workspace width. This reduction applies specifically to the reusable interaction workspace rather than the complete quantum resource requirement. Including the retained reward registers and the four-qubit return register, the complete dynamic implementation uses 15 logical qubits.

Under ideal simulation, the static and dynamic implementations produced the same set of 170 admissible three-step trajectories and the same return-grouping structure, with closely matched mean expected returns (\(5.9765 \pm 0.0709\) and \(5.9802 \pm 0.0675\), respectively). However, the mean total variation distance of \(0.2219\pm0.0143\) indicates that their finite-shot trajectory-probability distributions were not identical. These results suggest that dynamic-circuit qubit reuse preserves the admissible trajectory and return structure of the benchmark while changing its execution model, rather than reproducing an identical empirical trajectory distribution.

The framework also incorporates Grover-based trajectory search to identify the optimal quantum trajectories that yield the maximum return. Since the maximum return was known a priori from the reward structure, it was encoded directly in the oracle, and one Grover iteration, comprising one oracle invocation and one diffusion operation, was applied. Under ideal simulation, the marked trajectory T-151 became the most frequently observed trajectory, while the frequencies of other trajectories were suppressed. This demonstrates the integration of Grover amplitude amplification into the dynamic QMDP workflow as a unified quantum-native pipeline.

On the IBM Heron-class processor, the complete dynamic workflow was executable, but the measured trajectory distributions deviated substantially from their ideal counterparts. The observed distribution of outcomes across the reference trajectories indicates that the Grover search stage is sensitive to hardware noise and accumulated execution errors, and hardware-level amplification of the marked trajectory was not demonstrated. The individual contributions of different hardware error sources were not separately quantified. Although XY4 dynamical decoupling was enabled during hardware execution, no controlled comparison with and without this technique was performed. Consequently, the hardware experiments demonstrate implementation feasibility rather than hardware-level amplitude amplification, computational speedup, or quantum advantage.

\section{Conclusion}\label{sec:conclusion}
This paper presents a resource-efficient quantum reinforcement learning (QRL) architecture that integrates dynamic-circuit execution of a multi-step quantum Markov decision process (QMDP), quantum return evaluation, and Grover-based trajectory search. The primary contribution is an architectural reformulation of multi-step QMDP execution that reduces the interaction-workspace complexity from $O(T)$ to $O(1)$ with respect to the interaction horizon, without changing the underlying decision-making problem. Through mid-circuit measurement, reset, re-initialization, and sequential next-state propagation, the same state, action, next-state, and reward registers are reused across successive interactions. 
For the three-step benchmark, this architecture reduced the reusable interaction workspace from 21 to seven qubits, corresponding to a 66\% reduction, while the complete dynamic implementation required 15 logical qubits, including the additional registers for reward retention and return evaluation.

The architecture further integrates reversible quantum arithmetic for return evaluation and Grover-based trajectory search into a unified quantum workflow. Ideal simulations demonstrated that the dynamic implementation preserved the admissible trajectory set and corresponding returns of the static formulation, while the Grover stage demonstrated the amplification of the maximum-return trajectory. The complete dynamic workflow was also experimentally implemented on an IBM Heron-class processor, demonstrating its feasibility on current NISQ hardware. Nevertheless, the observed hardware distributional deviations and the absence of marked-trajectory amplification indicate that improving hardware reliability and search performance remains necessary.

The present evaluation is limited to a four-state, two-action, three-step proof-of-concept benchmark. Extending the framework to larger problems is left for future work, where state-preparation cost, retained quantum storage, circuit depth, dynamic-circuit latency, accumulated hardware errors, and trajectory-search complexity are likely to be the main constraints. Future research should investigate larger benchmarks, more efficient state-preparation schemes, reduced-depth return and oracle constructions, hardware-aware error-reduction strategies, and adaptive or maximum-finding procedures for cases in which the target return is not known a priori.

Overall, this study establishes a resource-oriented dynamic execution architecture for multi-step QMDPs, integrates quantum return evaluation and Grover-based trajectory search, and demonstrates the feasibility of executing the complete workflow on current quantum hardware. These results provide a foundation for further investigation of resource-efficient QRL architectures under the constraints of NISQ devices.

\funding{This work was supported by the Center of Innovations for Sustainable Quantum AI (JST Grant Number JPMJPF2221), JSPS KAKENHI Grant Number JP24K20843, and the Ministry of Education, Culture, Sports, Science and Technology (MEXT) of Japan. }

\dataavailability{The data supporting the findings of this study are available from the corresponding author(s) upon reasonable request.} 

\acknowledgments{We acknowledge the use of IBM Quantum services for this work. The views expressed are those of the authors, and do not reflect the official policy or position of IBM or the IBM Quantum team.}

\conflictsofinterest{The authors declare no conflicts of interest.} 

\appendixstart
\appendix
\section[\appendixname~\thesection]{Static QMDP Formulation as Baseline Implementation}
\label{appendixA}
This appendix presents the conventional static unrolled formulation of the QMDP, which serves as the baseline for comparison with the proposed resource-efficient dynamic-circuit implementation. The static formulation follows the design introduced in Ref.~\cite{thet2024}, in which each agent--environment interaction step is represented by an independent set of quantum registers.

In the static implementation, each interaction step \(t\in\{0,1,\ldots,T-1\}\) is realized using a distinct allocation of quantum registers: state register \texttt{qState$^{(t)}$}, action register \texttt{qAction$^{(t)}$}, next-state register \texttt{qNextState$^{(t)}$}, and reward register \texttt{qReward$^{(t)}$}. Therefore, the interaction-workspace requirement for a horizon of length \(T\) is \( N_{\mathrm{static,int}} = (2n_s+n_a+n_r)T.\)
For the benchmark considered in this study, with two state qubits, one action qubit, two next-state qubits, and two reward qubits per interaction, this gives \(N_{\mathrm{static,int}}=7T\). For \(T=3\), the static formulation requires 21 interaction-workspace qubits. Including the four-qubit return register, the complete static implementation uses 25 logical qubits.

\begin{figure}[H]
    \includegraphics[width=\textwidth]{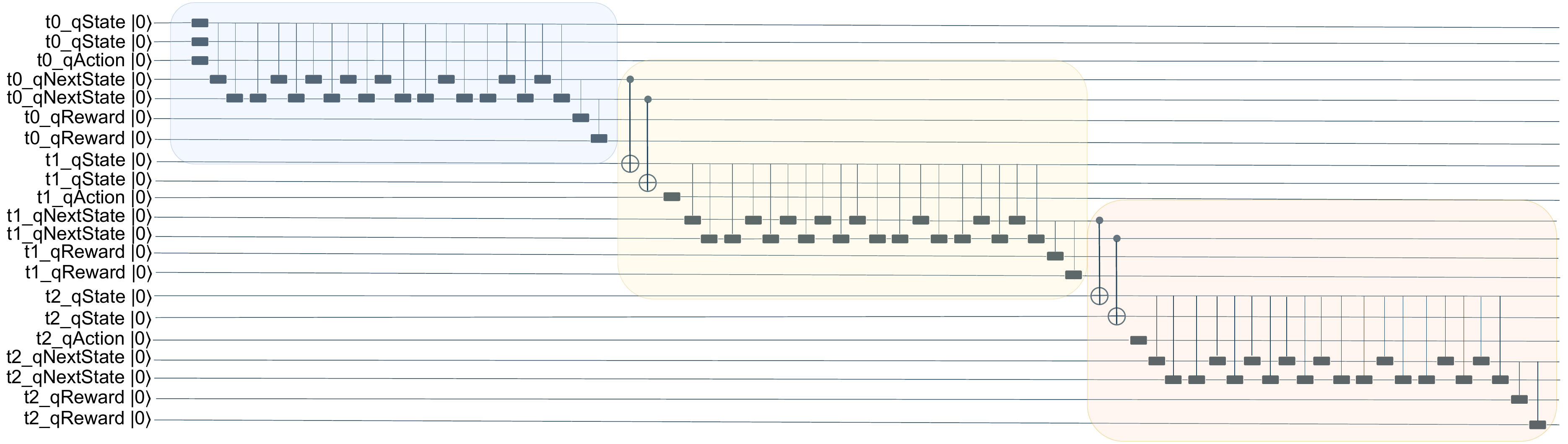}
    \vspace{-10pt}  
    \caption{Static QMDP circuit implementation of agent--environment interactions across three time steps (\(t=0,1,2\)). Each colored block corresponds to a single interaction step, with separate quantum registers allocated for the state (\texttt{qState}), action (\texttt{qAction}), next state (\texttt{qNextState}), and reward (\texttt{qReward}). CNOT gates propagate the next-state information between successive interaction blocks. Unlike the proposed dynamic-circuit implementation, the static formulation allocates distinct interaction registers at each time step, resulting in linear growth of the interaction-workspace width with the interaction horizon~\cite{thet2024}.}
    \label{fig:two_state_block_static}
\end{figure}

Figure~\ref{fig:two_state_block_static} illustrates the three-step static QMDP implementation. Each colored block corresponds to an independent interaction step. Unlike the dynamic-circuit implementation shown in Figure~\ref{fig:two_state_block}, the static formulation does not reuse the interaction registers; instead, separate interaction registers are retained across the unrolled horizon. Consequently, the interaction-workspace width increases linearly with the number of interaction steps.
For the three-step benchmark considered in this study, the static implementation produced the same set of 170 trajectories as the dynamic-circuit implementation under ideal simulation. Table~\ref{table:classical_trajectories} lists the complete trajectory set obtained from the static QMDP baseline.

Grover-based search is applied after the complete interaction horizon and quantum return calculation to increase the sampling probability of trajectories satisfying the maximum-return condition. 
For the three-step benchmark, the maximum return is \(g^\star=8\) (binary `1000'). The static Grover-based search identified the two maximum-return trajectories, T-143 and T-151, as the proposed dynamic implementation. For T-151, the corresponding action sequence is \(a_0\) at \(s_0\), followed by \(a_1\) at \(s_2\) and \(s_3\), providing a consistency check between the static and dynamic implementations.

\begin{table}[H]
\caption{Enumeration of the 170 quantum trajectories and corresponding trajectory indices obtained from the static QMDP implementation over three interaction steps. Each trajectory represents a state--action--next-state--reward sequence and is grouped according to its return. Detailed examples of the timestep-wise decomposition and qubit representation are provided in Tables~\ref{table:trajectories} and~\ref{table:qsampledistribution}, respectively.}
\label{table:classical_trajectories}

\begin{adjustwidth}{-\extralength}{-4cm}
\centering
\scriptsize
\setlength{\tabcolsep}{2pt}
\noindent
\begin{minipage}[t]{0.32\fulllength}
\vspace{0pt}
\centering
\renewcommand{\arraystretch}{0.45}
\begin{tabular}{@{}c@{\hskip -0.1cm}c@{}}
        \hline
        \hline
        \noalign{\vskip 2pt} 
        Trajectory no.  & Quantum trajectory \\
        \noalign{\vskip 1pt} 
        \hline
        \noalign{\vskip 2pt} 
        \multicolumn{2}{l}{\cellcolor{groupgray}\textbf{Return : 0001} \rule{0pt}{7pt}} \\
        \noalign{\vskip 2pt} 
             T-1 & ‘0001000000101010000000001' \\
             T-2 & ‘0001000000101010000000100' \\
             T-3 & ‘0001000000101011000000001' \\
             T-4 & ‘0001000000101011000000010' \\
             T-5 & ‘0001000010000000010101100' \\
             T-6 & ‘0001010100000001000000001' \\
             T-7 & ‘0001010100000001000000010' \\
             T-8 & ‘0001010110000001000000001' \\
             T-9 & ‘0001010110000001000000010' \\
        \multicolumn{2}{l}{\cellcolor{groupgray}\textbf{Return : 0010} \rule{0pt}{7pt}} \\
        \noalign{\vskip 2pt} 
             T-10 & ‘0010000000101010010101000' \\
             T-11 & ‘0010000000101010010101001' \\
             T-12 & ‘0010000000101010010101100' \\
             T-13 & ‘0010000000101010010101110' \\
             T-14 & ‘0010000001010100000000010' \\
             T-15 & ‘0010000010000000101010010' \\
             T-16 & ‘0010010100000000010101001' \\
             T-17 & ‘0010010100000000010101100' \\
             T-18 & ‘0010010100000000010101110' \\
             T-19 & ‘0010010100101010000000001' \\
             T-20 & ‘0010010100101010000000010' \\
             T-21 & ‘0010010100101010000000100' \\
             T-22 & ‘0010010100101011000000001' \\
             T-23 & ‘0010010100101011000000010' \\
             T-24 & ‘0010010100101011000000100' \\
             T-25 & ‘0010010110000000010101000' \\
             T-26 & ‘0010010110000000010101001' \\
             T-27 & ‘0010010110000000010101100' \\
             T-28 & ‘0010010110000000010101110' \\
             T-29 & ‘0010101000000001000000010' \\
        \multicolumn{2}{l}{\cellcolor{groupgray}\textbf{Return : 0011} \rule{0pt}{7pt}} \\
        \noalign{\vskip 2pt} 
             T-30 & ‘0011000000101011101010000' \\
             T-31 & ‘0011000000101011101010010' \\
             T-32 & ‘0011000001010101010101000' \\
             T-33 & ‘0011000001010101010101001' \\
             T-34 & ‘0011000001010101010101100' \\
             T-35 & ‘0011000001010101010101110' \\
             T-36 & ‘0011010100000000101010000' \\
             T-37 & ‘0011010100000000101010010' \\
             T-38 & ‘0011010100000000101010011' \\
             T-39 & ‘0011010100000000101010101' \\
             T-40 & ‘0011010100101010010101000' \\
             T-41 & ‘0011010100101010010101001' \\
             T-42 & ‘0011010100101010010101100' \\
             T-43 & ‘0011010100101010010101110' \\
             T-44 & ‘0011010110000000101010000' \\
             T-45 & ‘0011010110000000101010010' \\
             T-46 & ‘0011010110000000101010011' \\
             T-47 & ‘0011010110000000101010101' \\
             T-48 & ‘0011010111010100000000100' \\
             T-49 & ‘0011101000000000010101001' \\
             T-50 & ‘0011101000000000010101110' \\
             T-51 & ‘0011101010101010000000001' \\
             T-52 & ‘0011101010101010000000010' \\
             T-53 & ‘0011101010101011000000001' \\
             T-54 & ‘0011101010101011000000010' \\
        \multicolumn{2}{l}{\cellcolor{groupgray}\textbf{Return : 0100} \rule{0pt}{7pt}} \\
        \noalign{\vskip 2pt} 
             T-55 & ‘0100000001010100101010000' \\
             T-56 & ‘0100000001010100101010010' \\
             \noalign{\vskip 1pt} 
             \hline
             \hline
\end{tabular}
\end{minipage}
\hspace{-0.05\linewidth}
\begin{minipage}[t]{0.32\fulllength}
\vspace{0pt}
\centering
\renewcommand{\arraystretch}{0.495}
\begin{tabular}{@{}c@{\hskip -0.1cm}c@{}}
        \hline
        \hline
        \noalign{\vskip 2pt} 
        Trajectory no. & Quantum trajectory\\
        \noalign{\vskip 1pt} 
        \hline
        \noalign{\vskip 3pt} 
         T-57 & ‘0100000001010100101010011' \\
         T-58 & ‘0100000001010100101010101' \\
         T-59 & ‘0100010100101011101010000' \\
         T-60 & ‘0100010100101011101010010' \\
         T-61 & ‘0100010100101011101010011' \\
         T-62 & ‘0100010100101011101010101' \\
         T-63 & ‘0100010111010101010101000' \\
         T-64 & ‘0100010111010101010101001' \\
         T-65 & ‘0100010111010101010101100' \\
         T-66 & ‘0100010111010101010101110' \\
         T-67 & ‘0100101000000000101010011' \\
         T-68 & ‘0100101000000000101010101' \\
         T-69 & ‘0100101001010100000000001' \\
         T-70 & ‘0100101001010100000000010' \\
         T-71 & ‘0100101001010100000000100' \\
         T-72 & ‘0100101010101010010101000' \\
         T-73 & ‘0100101010101010010101001' \\
         T-74 & ‘0100101010101010010101100' \\
         T-75 & ‘0100101010101010010101110' \\
         T-76 & ‘0100111110101010000000010' \\
         T-77 & ‘0100111110101011000000001' \\
         T-78 & ‘0100111110101011000000010' \\
        \multicolumn{2}{l}{\cellcolor{groupgray}\textbf{Return : 0101} \rule{0pt}{7pt}} \\
        \noalign{\vskip 2pt} 
         T-79 & ‘0101000001010100111111011' \\
         T-80 & ‘0101000001010100111111101' \\
         T-81 & ‘0101000001010100111111110' \\
         T-82 & ‘0101000001010100111111111' \\
         T-83 & ‘0101010111010100101010010' \\
         T-84 & ‘0101010111010100101010011' \\
         T-85 & ‘0101010111010100101010101' \\
         T-86 & ‘0101101001010101010101000' \\
         T-87 & ‘0101101001010101010101001' \\
         T-88 & ‘0101101001010101010101100' \\
         T-89 & ‘0101101001010101010101110' \\
         T-90 & ‘0101101010101011101010010' \\
         T-91 & ‘0101101010101011101010011' \\
         T-92 & ‘0101101010101011101010101' \\
         T-93 & ‘0101111110101010010101000' \\
         T-94 & ‘0101111110101010010101001' \\
         T-95 & ‘0101111110101010010101100' \\
         T-96 & ‘0101111111010100000000001' \\
         T-97 & ‘0101111111010100000000010' \\
         T-98 & ‘0101111111010100000000100' \\
        \multicolumn{2}{l}{\cellcolor{groupgray}\textbf{Return : 0110} \rule{0pt}{7pt}} \\
        \noalign{\vskip 2pt} 
         T-99 & ‘0110010111010100111111011' \\
         T-100 & ‘0110010111010100111111101' \\
         T-101 & ‘0110010111010100111111110' \\
         T-102 & ‘0110010111010100111111111' \\
         T-103 & ‘0110101001010100101010000' \\
         T-104 & ‘0110101001010100101010010' \\
         T-105 & ‘0110101001010100101010011' \\
         T-106 & ‘0110101001010100101010101' \\
         T-107 & ‘0110101001111111010101000' \\
         T-108 & ‘0110101001111111010101001' \\
         T-109 & ‘0110101001111111010101100' \\
         T-110 & ‘0110111110101011101010000' \\
         T-111 & ‘0110111110101011101010011' \\
         T-112 & ‘0110111110101011101010101' \\
         T-113 & ‘0110111111010101010101000' \\
         T-114 & ‘0110111111010101010101001' \\
         \noalign{\vskip 1pt} 
         \hline
         \hline
\end{tabular}
\end{minipage}
\hspace{-0.05\linewidth}
\begin{minipage}[t]{0.32\fulllength}
\vspace{0pt}
\centering
\renewcommand{\arraystretch}{0.51}
\begin{tabular}{@{}c@{\hskip -0.1cm}c@{}}
        \hline
        \hline
        \noalign{\vskip 2pt} 
        Trajectory no. & Quantum trajectory\\
        \noalign{\vskip 1pt} 
        \hline
        \noalign{\vskip 3pt} 
         T-115 & ‘0110111111010101010101100' \\
        \multicolumn{2}{l}{\cellcolor{groupgray}\textbf{Return : 0111} \rule{0pt}{7pt}} \\
        \noalign{\vskip 2pt} 
         T-116 & ‘0111101001010100111111011' \\
         T-117 & ‘0111101001010100111111101' \\
         T-118 & ‘0111101001010100111111110' \\
         T-119 & ‘0111101001010100111111111' \\
         T-120 & ‘0111101001111111101010000' \\
         T-121 & ‘0111101001111111101010010' \\
         T-122 & ‘0111101001111111101010011' \\
         T-123 & ‘0111101001111111101010101' \\
         T-124 & ‘0111111101111111010101000' \\
         T-125 & ‘0111111101111111010101001' \\
         T-126 & ‘0111111101111111010101100' \\
         T-127 & ‘0111111111010100101010000' \\
         T-128 & ‘0111111111010100101010010' \\
         T-129 & ‘0111111111010100101010011' \\
         T-130 & ‘0111111111010100101010101' \\
         T-131 & ‘0111111111111111010101000' \\
         T-132 & ‘0111111111111111010101001' \\
         T-133 & ‘0111111111111111010101100' \\
         T-134 & ‘0111111111111111010101110' \\
        \multicolumn{2}{l}{\cellcolor{groupgray}\textbf{Return : 1000} \rule{0pt}{7pt}} \\
        \noalign{\vskip 2pt} 
         T-135 & ‘1000101001111110111111011' \\
         T-136 & ‘1000101001111110111111101' \\
         T-137 & ‘1000101001111110111111110' \\
         T-138 & ‘1000101001111110111111111' \\
         T-139 & ‘1000101001111111111111011' \\
         T-140 & ‘1000101001111111111111101' \\
         T-141 & ‘1000101001111111111111110' \\
         T-142 & ‘1000101001111111111111111' \\
         T-143 & ‘1000111101111111101010000' \\
         T-144 & ‘1000111101111111101010010' \\
         T-145 & ‘1000111101111111101010011' \\
         T-146 & ‘1000111101111111101010101' \\
         T-147 & ‘1000111111010100111111011' \\
         T-148 & ‘1000111111010100111111101' \\
         T-149 & ‘1000111111010100111111110' \\
         T-150 & ‘1000111111010100111111111' \\
         T-151 & ‘1000111111111111101010000' \\
         T-152 & ‘1000111111111111101010010' \\
         T-153 & ‘1000111111111111101010011' \\
         T-154 & ‘1000111111111111101010101' \\
        \multicolumn{2}{l}{\cellcolor{groupgray}\textbf{Return : 1001} \rule{0pt}{7pt}} \\
        \noalign{\vskip 2pt} 
         T-155 & ‘1001111101111110111111011' \\
         T-156 & ‘1001111101111110111111101' \\
         T-157 & ‘1001111101111110111111110' \\
         T-158 & ‘1001111101111110111111111' \\
         T-159 & ‘1001111101111111111111011' \\
         T-160 & ‘1001111101111111111111101' \\
         T-161 & ‘1001111101111111111111110' \\
         T-162 & ‘1001111101111111111111111' \\
         T-163 & ‘1001111111111110111111011' \\
         T-164 & ‘1001111111111110111111101' \\
         T-165 & ‘1001111111111110111111110' \\
         T-166 & ‘1001111111111110111111111' \\
         T-167 & ‘1001111111111111111111011' \\
         T-168 & ‘1001111111111111111111101' \\
         T-169 & ‘1001111111111111111111110' \\
         T-170 & ‘1001111111111111111111111' \\
         \noalign{\vskip 1pt} 
         \hline
         \hline
\end{tabular}
\end{minipage}
\end{adjustwidth}
\end{table}


\reftitle{References}
\bibliography{references}


%


\end{document}